\documentclass{emulateapj2}

\usepackage{graphicx} 
\usepackage{longtable} 
\usepackage{rotating}
\usepackage{tabularx}
\usepackage{hyperref}
\usepackage{easylist}
\usepackage{amsmath}
\usepackage{bm}
\usepackage{array} % for extrarowheight
\usepackage{amssymb,amsmath}

\newcommand{\numax}{\mbox{$\nu_{\rm max}$}}
\newcommand{\Dnu}{\mbox{$\Delta \nu$}}

\newcommand{\muHz}{\mbox{$\mu$Hz}}
\newcommand{\kep}{\mbox{\textit{Kepler}}}
\newcommand{\teff}{\mbox{$T_{\rm eff}$}}
\newcommand{\logg}{\mbox{$\log g$}}
\newcommand{\feh}{\mbox{$\rm{[Fe/H]}$}}

\newcommand{\rsun}{\mbox{$R_{\sun}$}}

\newcommand{\nstars}{\mbox{2200}}
\newcommand{\ndwarfs}{\mbox{440}}
\newcommand{\ngiants}{\mbox{1800}}

\newcommand\blfootnote[1]{%
  \begingroup
  \renewcommand\thefootnote{}\footnote{#1}%
  \addtocounter{footnote}{-1}%
  \endgroup
}

\include{journals}

\slugcomment{accepted for publication in ApJ}

\shorttitle{Asteroseismology and Gaia: Testing Scaling Relations}
\shortauthors{Daniel Huber et al.}

\begin{document}

\title{Asteroseismology and Gaia: Testing Scaling Relations Using 2200 Kepler Stars with TGAS Parallaxes}

\author{
Daniel Huber\altaffilmark{1,2,3,4}, 
Joel Zinn\altaffilmark{5}, 
Mathias Bojsen-Hansen\altaffilmark{4}, 
Marc Pinsonneault\altaffilmark{5}, 
Christian Sahlholdt\altaffilmark{4}, 
Aldo Serenelli\altaffilmark{6}, 
Victor Silva Aguirre\altaffilmark{4}, 
Keivan Stassun\altaffilmark{7,8}, 
Dennis Stello\altaffilmark{9,2,4}, 
Jamie Tayar\altaffilmark{5}, 
Fabienne Bastien\altaffilmark{10,11,18}, 
Timothy R.~Bedding\altaffilmark{2,4}, 
Lars A.~Buchhave\altaffilmark{12}, 
William J.~Chaplin\altaffilmark{13,4}, 
Guy R.~Davies\altaffilmark{13,4}, 
Rafael A.~Garc\'\i a\altaffilmark{14}, 
David W.~Latham\altaffilmark{15}, 
Savita Mathur\altaffilmark{16}, 
Benoit Mosser\altaffilmark{17} and Sanjib Sharma\altaffilmark{2}  
}
\altaffiltext{1}{Institute for Astronomy, University of Hawai`i, 2680 Woodlawn Drive, Honolulu, HI 96822, USA; \mbox{huberd@hawaii.edu}}
\altaffiltext{2}{Sydney Institute for Astronomy (SIfA), School of Physics, University of Sydney, NSW 2006, Australia}
\altaffiltext{3}{SETI Institute, 189 Bernardo Avenue, Mountain View, CA 94043, USA}
\altaffiltext{4}{Stellar Astrophysics Centre, Department of Physics and Astronomy, Aarhus University, Ny Munkegade 120, DK-8000 Aarhus C, Denmark}
\altaffiltext{5}{Department of Astronomy, The Ohio State University, Columbus, OH 43210, USA}
\altaffiltext{6}{Institute of Space Sciences (IEEC-CSIC), Campus UAB, Carrer de Can Magrans S/N, 08193, Barcelona, Spain}
\altaffiltext{7}{Vanderbilt University, Department of Physics and Astronomy, 6301 Stevenson Center Lane, Nashville, TN 37235, USA}
\altaffiltext{8}{Fisk University, Department of Physics, 1000 17th Avenue N., Nashville, TN 37208, USA}
\altaffiltext{9}{School of Physics, University of New South Wales, NSW 2052, Australia}
\altaffiltext{10}{Department of Astronomy and Astrophysics, The Pennsylvania State University, 525 Davey Laboratory, University Park, PA 16802, USA}
\altaffiltext{11}{Center for Exoplanets and Habitable Worlds, The Pennsylvania State University, 525 Davey Laboratory, University Park, PA 16802, USA}
\altaffiltext{12}{Centre for Star and Planet Formation, Natural History Museum of Denmark \& Niels Bohr Institute, University of Copenhagen, Oster Voldgade 5-7, DK-1350 Copenhagen K, Denmark}
\altaffiltext{13}{School of Physics and Astronomy, University of Birmingham, Birmingham B15 2TT, UK}

\begin{abstract}
We present a comparison of parallaxes and radii from asteroseismology and Gaia DR1 (TGAS) for \nstars\ \kep\ stars spanning from the main sequence to the red giant branch. We show that previously identified offsets between TGAS parallaxes and distances derived from asteroseismology and eclipsing binaries have likely been overestimated for parallaxes $\lesssim 5-10$\,mas ($\approx$\,90--98\,\% of the TGAS sample). The observed differences in our sample can furthermore be partially compensated by adopting a hotter $\teff$ scale (such as the infrared flux method) instead of spectroscopic temperatures for dwarfs and subgiants. Residual systematic differences are at the $\approx$\,2\,\% level in parallax across three orders of magnitude. We use TGAS parallaxes to empirically demonstrate that asteroseismic radii are accurate to $\approx$\,5\,\% or better for stars between $\approx 0.8-8\,\rsun$. We find no significant offset for main-sequence ($\lesssim 1.5\rsun$) and low-luminosity RGB stars ($\approx$\,3--8\rsun), but seismic radii appear to be systematically underestimated by $\approx$\,5\% for subgiants ($\approx$1.5--3\rsun). We find no systematic errors as a function of metallicity between $\feh \approx -0.8$ to $+0.4$\,dex, and show tentative evidence that corrections to the scaling relation for the large frequency separation ($\Dnu$) improve the agreement with TGAS for RGB stars. Finally, we demonstrate that beyond $\approx 3$\,kpc asteroseismology will provide more precise distances than end-of-mission Gaia data, highlighting the synergy and complementary nature of Gaia and asteroseismology for studying galactic stellar populations. 
%Finally, we release the open-source software tool \texttt{isoclassify}\footnote{\url{https://github.com/danxhuber/isoclassify}}, which efficiently derives stellar properties given any combination of photometric, spectroscopic, asteroseismic, and astrometric observables.
\end{abstract}

%\tableofcontents

\section{Introduction}

Over the past decade asteroseismology has emerged as an important method to systematically determine fundamental properties of stars. For example, asteroseismology has been used to determine precise radii, masses and ages of exoplanet host stars \citep{cd10,huber13b,silva15}, calibrate spectroscopic surface gravities \citep{brewer15,petigura15b,wang16}, and study masses and ages of galactic stellar populations \citep{miglio09,casagrande14,mathur16,anders17}. Due to the wealth of data from space-based missions \citep{chaplin13a} and the complexity of modeling oscillation frequencies for evolved stars \citep[e.g.][]{dimauro11}, most studies have relied on global asteroseismic observables and scaling relations to derive fundamental stellar properties. Testing the validity of these scaling relations has become one of the most active topics in asteroseismology.
% ridiculous hack to prevent emulateapj to screw up and put introduction on page 2
\blfootnote{$^{14}$Laboratoire AIM, CEA/DRF-CNRS, Universit\'e Paris 7 Diderot, IRFU/SAp, Centre de Saclay, 91191, Gif-sur-Yvette, France}
\blfootnote{$^{15}$ Harvard-Smithsonian Center for Astrophysics, 60 Garden Street, Cambridge, Massachusetts 02138, USA}
\blfootnote{$^{16}$ Space Science Institute, 4750 Walnut street Suite 205, Boulder, CO 80301, USA}
\blfootnote{$^{17}$ LESIA, Observatoire de Paris, PSL Research University, CNRS, Universit\'e Pierre et Marie Curie, Universit\'e Paris Diderot, 92195 Meudon, France}
\blfootnote{$^{18}$ Hubble Fellow}
\setcounter{footnote}{18}

Empirical tests have so far included interferometry \citep{huber12, white13}, Hipparcos parallaxes \citep{silva12}, eclipsing binaries \citep{frandsen13,huber14b,gaulme16} and open clusters \citep{miglio11, miglio16, stello16b}. These tests have indicated that scaling relations are accurate to within $\approx$\,5\% in radius for main-sequence stars, while larger discrepancies have been identified for giants. In particular, \citet{gaulme16} reported a systematic overestimation of $\approx$\,5\% in radius and $\approx$\,15\% in mass for red giants with $R \gtrsim 8\rsun $, based on a comparison with dynamical properties derived from double-lined eclipsing binaries. Semi-empirical tests using halo stars have furthermore indicated that masses from scaling relations are significantly overestimated compared to expectation values for luminous metal-poor ($\feh < -1$) giants \citep{epstein14}. Population synthesis models also suggest that the observed mass distributions are shifted towards higher masses compared to predictions \citep{sharma16,sharma17}.

Theoretical work has motivated corrections to scaling relations, for example by comparing the large frequency separation (\Dnu) calculated from individual frequencies with model densities \citep{stello09c,white11,guggenberger16,sharma16} or an extension of the asymptotic relation \citep{mosser13}. A consistent result is that $\Dnu$ scaling relation corrections should depend on \teff, evolutionary state and metallicity. However, it is as of yet unclear whether these corrections are more important for red-giant branch or red clump stars \citep{miglio11,sharma16}. Additionally, uncertainties in modeling the driving and damping of oscillations typically prevent theoretical tests of the $\numax$ scaling relation, although some studies have shown encouraging results \citep{belkacem11}.% towards that goal.

Despite these efforts, the validity of scaling relations as a function of metallicity and evolutionary state is poorly tested. Gaia parallaxes are dramatically improving this situation by providing a large set of distances for asteroseismic samples observed by \kep. Initial comparisons with Gaia DR1 (TGAS) using published asteroseismic distances indicated good agreement for 20 nearby dwarfs \citep{deridder16}. However, subsequent work by \citet{silva17} using a sample of $\approx$\,60 nearby dwarfs revealed a systematic offset between TGAS and asteroseismology, in agreement with results from eclipsing binaries by \citet{stassun16b} and ground-based parallaxes for dwarfs at $<$\,25\,pc \citep{jao16}. \citet{deridder16} also found a discrepancy for $\approx$\,900 giants, which was explained as a systematic bias in TGAS parallaxes based on a comparison to red clump stars \citep{davies17}. Combined with the absence of offsets for distant Cepheids \citep{sesar16}, these results have been interpreted as evidence for a distance-dependent systematic error in TGAS parallaxes as large as $\approx$\,0.39\,mas down to $\pi \approx$\,2\,mas ($\approx$\,20\%).

%First comparisons to Gaia DR1 (TGAS) using published asteroseismic distances indicated good agreement for dwarfs but a discrepancy for giants \citep{deridder16}, which was later explained as a systematic error in TGAS parallaxes based on a comparison to red clump stars \citep{davies17}. 

\begin{figure}
\begin{center}
\resizebox{\hsize}{!}{\includegraphics{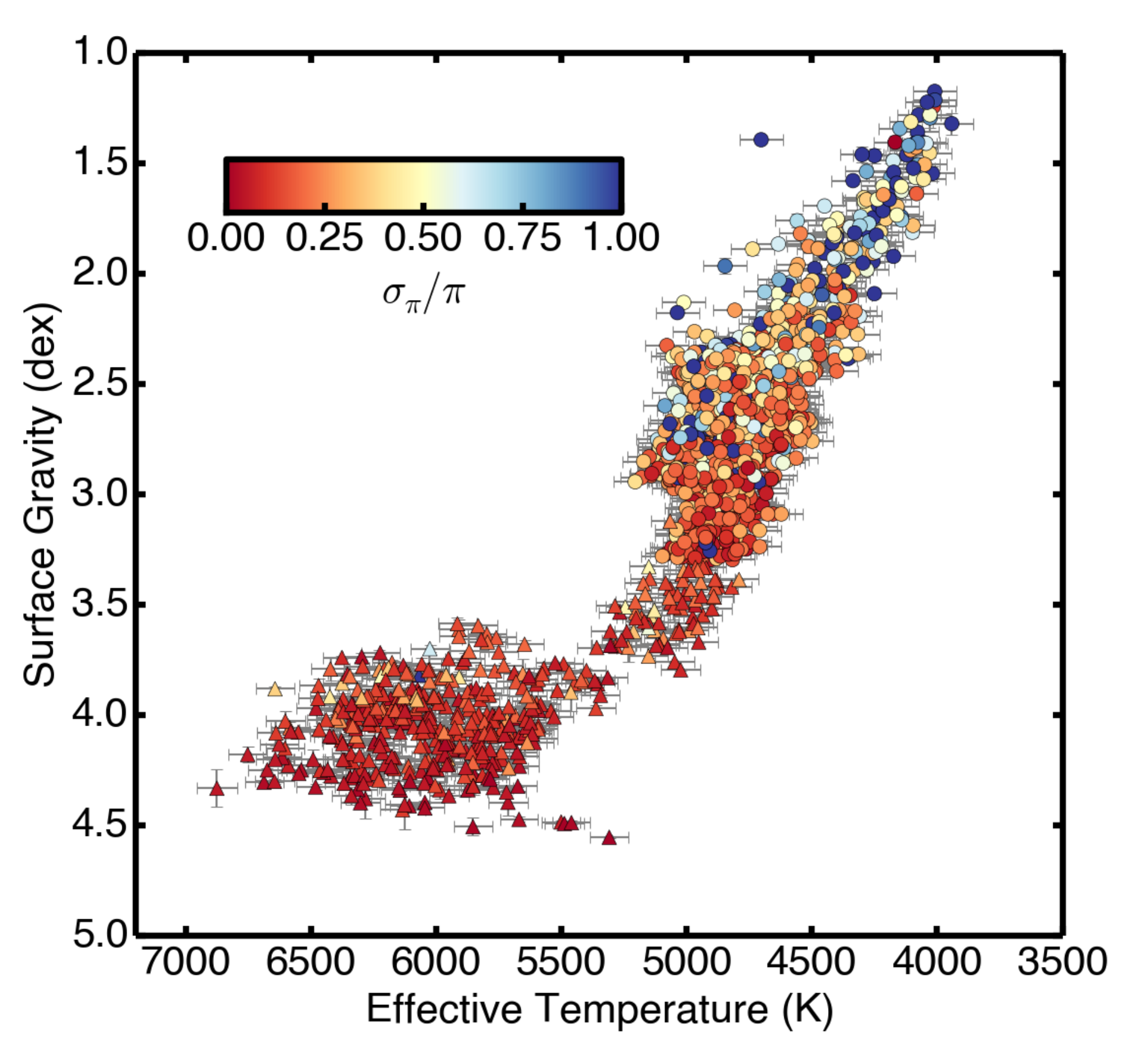}}
\caption{Surface gravity versus effective temperature for $\approx$\,\ngiants\ red giants and $\approx$\,\ndwarfs\ dwarfs and subgiants with TGAS parallaxes and detected oscillations from \kep. The fractional TGAS parallax precision is color-coded (the color scale is capped at 1 for clarity). Triangles and circles show stars with \teff\ and \feh\ from optical (SPC, $\logg \gtrsim 3.4$) and infrared (APOGEE, $\logg \lesssim 3.4$) spectroscopy, respectively.}
\label{fig:sample}
\end{center}
\end{figure}

Here we use TGAS parallaxes for a large sample of \nstars\ \kep\ stars to revisit the comparison between TGAS and asteroseismology, and to test asteroseismic scaling relations. Unlike previous studies our sample has continuous coverage from the main sequence to the red-giant branch, allowing us to compare TGAS and asteroseismology over a range of distances and evolutionary states. A companion paper describes an investigation of correlated spatial differences between TGAS and the asteroseismic \kep\ sample (Zinn et al., in prep).

\section{Target Sample}

Our sample consists of dwarfs, subgiants, and red giants from the APOGEE-Kepler Asteroseismic Science Consortium \citep[APOKASC,][]{pinsonneault14}, supplemented with seismic detections using \kep\ short-cadence data from \citet{chaplin14} and \citet{huber13}. Effective temperatures and metallicities for dwarfs and subgiants were obtained from an SPC analysis of optical high-resolution spectra obtained with the TRES spectrograph at the F.\ L.\ Whipple Observatory \citep{buchhave12, buchhave14}. The SPC analysis was performed with externally constrained asteroseismic $\logg$ values, which prevents degeneracies between $\teff$, $\logg$ and $\feh$ \citep{torres12,huber13}. For giants, we adopted ASPCAP parameters from SDSS DR13 \citep{holtzman15, sdssdr13}. We furthermore collected asteroseismic parameters \numax\ and \Dnu\ from a reanalysis of the \citet{chaplin14} sample using all available \kep\ data for dwarfs and subgiants (Serenelli et al., in prep), and version 3.6.5 of the APOKASC catalog for giants (Pinsonneault et al., in prep). We adopted values from the SYD pipeline \citep{huber09}, but note that differences between asteroseismic pipelines do not affect the conclusions in this paper (see also Section \ref{sec:seismovalidation}). Finally, we collected $griz$ photometry from the Kepler Input Catalog \citep[KIC,][]{brown11}, corrected to the SDSS scale following \citet{pinsonneault11}, 2MASS $JHK$, Tycho $B_{T}V_{T}$, and TGAS parallaxes \citep{gaia1,gaia2,lindegren16} for each star. Our final sample contains $\approx$\,\ndwarfs\ dwarfs and subgiants as well as over \ngiants\ red giants with asteroseismic parameters, broadband photometry, and parallaxes. Table \ref{tab:observables} lists all observables used in this study. Unless otherwise noted, all results in this paper are based on the combination of \teff\ and \feh\ from APOGEE and SPC, as described above.
%We added a systematic error of 0.3\,mas in quadrature to all parallax measurements.

Figure \ref{fig:sample} shows the sample in a \teff-\logg\ diagram, with the fractional TGAS parallax uncertainty color-coded. As expected the fractional parallax uncertainty is a strong function of distance and hence evolutionary state: dwarfs and subgiants have a typical fractional uncerainty of $\approx$\,5\%, increasing to $\approx$\,10\% for subgiants and $\approx$\,50\% for red clump stars. Compared to Hipparcos, this sample increases the number of asteroseismic \kep\ stars with parallaxes by a factor of $\approx20$.

\section{Methodology}

\subsection{Direct Method}

Scaling relations for solar-like oscillations are based on the global asteroseismic observables $\numax$, the frequency of maximum power, and $\Dnu$, the average separation of oscillation modes with the same spherical degree and consecutive radial order. The relations are defined as follows \citep{kb95}:

\begin{equation}
\Delta\nu \propto \left(\frac{M}{R^3}\right)^{1/2} \: ,
\label{equ:delnu}
\end{equation}

\begin{equation}
\nu_{\rm max} \propto \frac{M}{R^2 \sqrt{T_{\rm eff}}} \: .
\label{equ:numax}
\end{equation}

Equations (\ref{equ:delnu}) and (\ref{equ:numax}) can be rearranged to calculate radius as follows:

\begin{equation}
\frac{R}{\mathrm{R}_\odot} \approx \left(\frac{\nu_\mathrm{max}}{\nu_\mathrm{max,\odot}}\right)\left(\frac{\Delta\nu}{\Delta\nu_\odot}\right)^{-2}\left(\frac{T_\mathrm{eff}}{\mathrm{T_{eff,\odot}}}\right)^{1/2}.
\label{equ:rad}
\end{equation}

We used $\numax_{\sun}=3090$\,\muHz\ and $\Dnu_{\sun}=135.1$\,\muHz, the solar reference values for the SYD pipeline \citep{huber11}. Corrections for the \Dnu\ scaling relation (see Section 1) were calculated using \texttt{asfgrid} \citep{sharma16}\footnote{\url{http://www.physics.usyd.edu.au/k2gap/Asfgrid/}}. To calculate asteroseismic distances, we combined \teff\ with the radius from equation (\ref{equ:rad}) to calculate luminosity, and then used the 2MASS $K$-band magnitude with bolometric corrections derived by linearly interpolating  \teff, \logg, \feh\ and $A_{V}$ in the MIST/C3K grid (Conroy et al., in prep\footnote{\url{http://waps.cfa.harvard.edu/MIST/model_grids.html}}). To estimate $A_{V}$ we used the 3D reddening map by \citet{green15}, as implemented in the \texttt{mwdust} package by \citet{bovy16}. The derived distances, extinction values and bolometric corrections were iterated until convergence. 

Parallaxes can also be used to calculate luminosities (and hence radii), which can be compared to asteroseismic radii. To convert parallaxes into distances we used an exponentially decreasing volume density prior with a length scale of 1.35\,kpc \citep{bailer15,astraatmadja16}. In practice, we implemented a Monte-Carlo method by sampling distances following the distance posterior distribution. For each distance sample, we calculated reddening given the 3D dust map, and combined this with samples for the apparent magnitude and \teff\ (drawn from a random normal distribution with a standard deviation corresponding to the 1-$\sigma$ uncertainties) to calculate radii. The adopted bolometric corrections and \teff\ values were identical to the ones used for the calculation of asteroseismic radii described above.

The resulting distributions were used to calculate the mode and 1-$\sigma$ confidence interval for radii derived from each Gaia parallax. We did not implement a more complex prior (e.g. based on synthetic stellar population) due to the difficulty of reproducing the selection function of our sample, but note that the results in this paper do not heavily depend on the choice of distance prior.

\begin{figure*}
\begin{center}
\resizebox{\hsize}{!}{\includegraphics{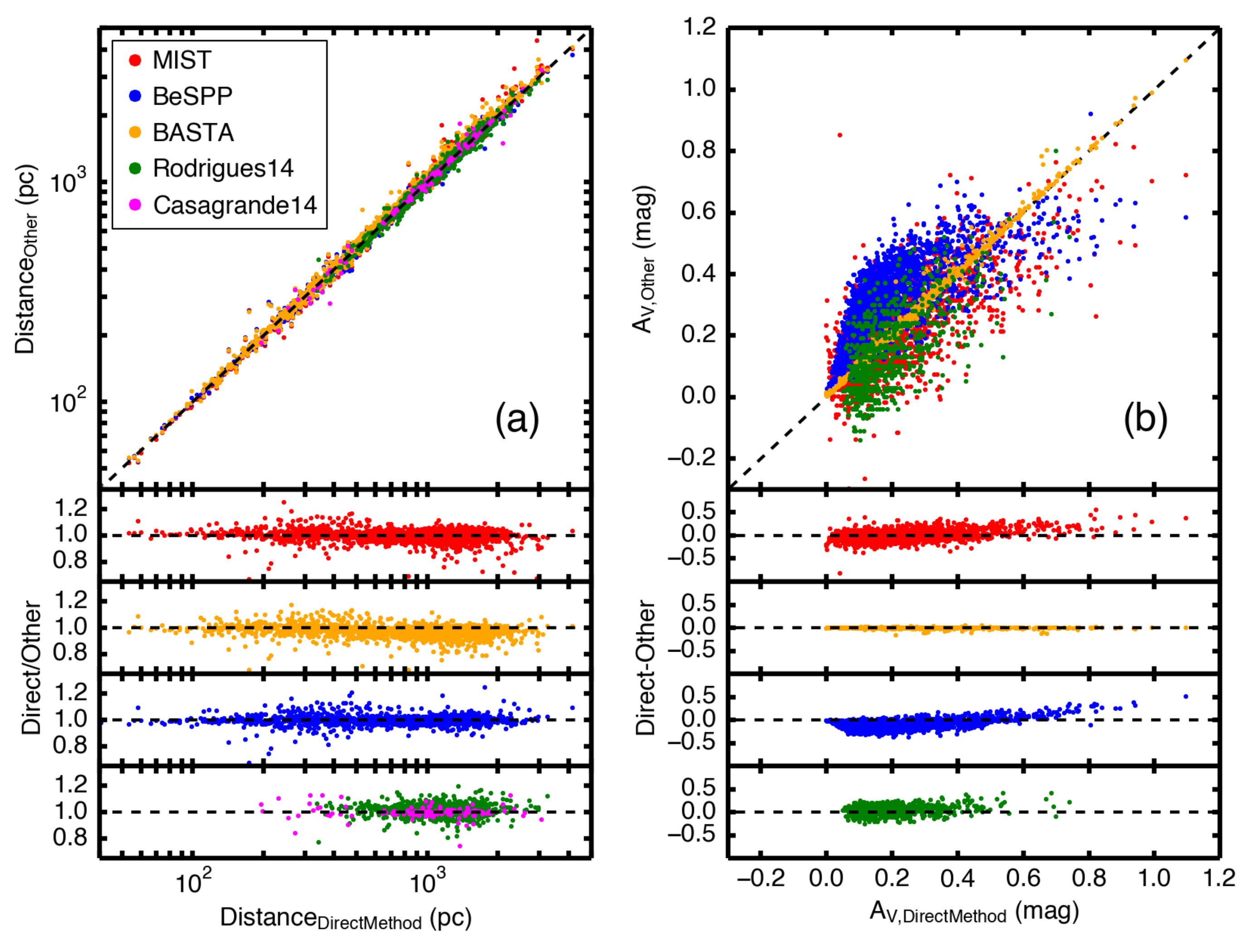}}
\caption{Panel a: Comparison of asteroseismic distances from the direct method with grid-modelled distances derived using MIST (this work), BeSPP \citep{serenelli13}, BASTA \citep{silva15}, as well as values from the SAGA survey \citep{casagrande14} and \citet{rodrigues14}. The top panel shows the 1:1 relation, and the bottom panels show residuals. Panel b: Comparison of extinction values from the 3D map by \citet{green15} (as applied in the direct method) with values derived by combining asteroseismology, spectroscopy and photometry (MIST and \citet{rodrigues14}) and the extinction model by \citet{amores05} (as applied in BeSPP). Colors mark the same datasets as in panel a. Note that BASTA uses the same reddening map as the direct method in this work.}
\label{fig:discomp}
\end{center}
\end{figure*}

\subsection{Grid Modeling}

The ``direct method'' for determining asteroseismic distances described in the previous section has the disadvantage that it relies on a reddening map, which may contain systematic errors. We therefore calculated a second set of asteroseismic distances and TGAS radii using isochrones and synthetic photometry, which allows reddening to be treated as a free parameter. We used isochrones from the MIST database \citep{choi16,paxton11,paxton13,paxton15} to calculate a grid ranging in age from 0.5 to 14\,Gyr with a stepsize of 0.25\,Gyr and in metallicity from $-2$ to $+0.4$\,dex in stepsizes of 0.02\,dex. Interpolation was performed along equal evolutionary points in age and metallicity \citep{dotter16}. For each model we saved synthetic photometry in 2MASS $JHK$, Tycho $B_{T}V_{T}$, and Sloan $griz$, and calculated reddened photometry in each passband for a given $V$-band extinction $A_{V}$ by interpolating the \citet{cardelli89} extinction law. Asteroseismic $\numax$ and $\Dnu$ values for each model were calculated using Equations (\ref{equ:delnu}) and (\ref{equ:numax}), both with and without the $\Dnu$ scaling relation corrections by \citet{sharma16}.

To infer model parameters we followed the method by \citet{serenelli13} to integrate over all isochrone points to derive posterior distributions given a set of likelihoods and priors. Specifically, given any combination of a set of observables $x=\{B_{T}-V_{T},g-r,r-i,i-z,J-H,H-K,\pi,\teff,\feh,\numax,\Dnu\}$ and model parameters $y=\{\rm age,\ [Fe/H],\ mass, A_{V}\}$, the posterior probability is:

\begin{equation}
p(y|x) \propto p(y)p(x|y) \propto  p(y)\prod_i \exp{\left(-\frac{(x_i-x_i(y))^2}{2 \sigma_{x,i}^2}\right)} \: .
\end{equation}

The likelihood function for $\pi$ was calculated as \citep[e.g.][]{bailer15}:

\begin{equation}
p(\pi|d) \propto \exp{\left[ - \frac{1}{2\sigma_{\pi}^{2}} \left(\pi - \frac{1}{d}\right)^{2}    \right]} \: ,
\end{equation}

\noindent
where $d$ is the model distance calculated given an absolute magnitude and $A_{V}$ for each model, as well as the observed $K$-band magnitude. Probability distribution functions for each stellar parameter were then obtained by weighting $p(y|x)$ by the volume which each isochrone point encompasses in mass, age, metallicity, and $A_{V}$, and integrating the resulting distribution along a given stellar parameter \citep[see appendix A of][]{casagrande11}. For ease of computation, the integration was performed only for models within 4-$\sigma$ of the constraints set by the observables.

To calculate asteroseismic distances we used as input the spectroscopic $\teff$ and $\feh$, asteroseismic $\numax$ and $\Dnu$, $B_{T}V_{T}griJHK$ photometry, and a flat prior in age, resulting in posterior distributions for all stellar parameters as well as extinction and distance. To calculate TGAS radii we replaced the asteroseismic observables with the TGAS parallax $\pi$, using a flat age prior and the same distance prior as adopted in the previous section.

\subsection{Validation of Seismic Distances and Gaia Radii}

\subsubsection{Asteroseismic Parameters}
\label{sec:seismovalidation}

Comparisons of different methods to measure asteroseismic parameters have yielded broadly good agreement \citep{hekker11,verner11,hekker12b}. The median scatter between the five methods in the APOKASC catalog \citep[see ][]{pinsonneault14} is 0.5\% in \Dnu\ and 1\% in \numax, which we added in quadrature to the formal uncertainties from the SYD pipeline (see Table \ref{tab:observables}) for the analysis described in the previous section. 

To test the influence of systematic errors, we compared our asteroseismic distances calculated using the direct and grid modeling method in Figure \ref{fig:discomp}a. The agreement is excellent, with median offset of $0.2$\%\ and scatter of 2.6\%. To test a variety of systematic errors that could enter the asteroseismic distance calculation, we  compared our distances from the direct method with distances calculated using the Bellaterra Stellar Properties Pipeline \citep[BeSPP,][]{serenelli13}, the BAyesian STellar Algorithm \citep[BASTA,][]{silva15}, as well as to literature values from the Stromgren Survey for Asteroseismology and Galactic Archeology \citep[SAGA,][]{casagrande14} and \citet{rodrigues14}. 
Three of these methods (BeSPP, BASTA, SAGA) used asteroseismic input values from the same pipeline but different isochrone grids, and one method used a different asteroseismic input values and isochrone models \citep{rodrigues14}. The median offsets are $\approx$\,0.2\% for BeSPP, 2.3\% for BASTA, 0.1\% for SAGA and 1.8\% for \citet{rodrigues14}, with no strong systematic trends as a function of distance (see bottom panel of Figure \ref{fig:discomp}a). We thus conclude that systematic differences between asteroseismic methods to calculate distances are of the order of a few percent.

\subsubsection{Extinction}

Asteroseismic distances rely on extinction corrections, which can introduce systematic errors. Figure \ref{fig:discomp}b compares the extinction measured using our grid-based method with the reddening map by \citet{green15}, as applied in our direct method. We also show extinctions from \citet{rodrigues14}, which were derived in a similar manner to the grid-modeling estimates presented here, and values from the model by \citet{amores05}, as applied by the BeSPP pipeline. The estimates agree well for $A_{V}\lesssim 0.5$\,mag, with a slight systematic overestimation by up to 0.2\,mag of the \citet{green15} reddening map for $A_{V}\gtrsim 0.5$\,mag. This comparison demonstrates that the combination of spectroscopy, asteroseismology and Gaia has strong potential for constructing empirical 3D reddening maps, in particular when combined with asteroseismic detections in different regions of the galaxy as provided by CoRoT \citep{hekker09} and K2 \citep{stello17}.

We note that the slight bias for high extinction in Figure \ref{fig:discomp}b has only a small effect on Figure \ref{fig:discomp}a, since the sample is dominated by stars with low extinction. Additionally, a systematic shift of 0.2\,mag in $A_{V}$ corresponds to an error of 0.02\,mag in $A_{K}$, or $\lesssim$\,1\,\% in distance. Since distances from the direct method are the least model-dependent and more directly test the validity of scaling relations, we proceed with using these values for the remainder of the paper. We note that our main conclusions are independent of whether the direct method or the grid-modeling method is adopted.

\subsubsection{Bolometric Corrections}

\begin{figure}
\begin{center}
\resizebox{\hsize}{!}{\includegraphics{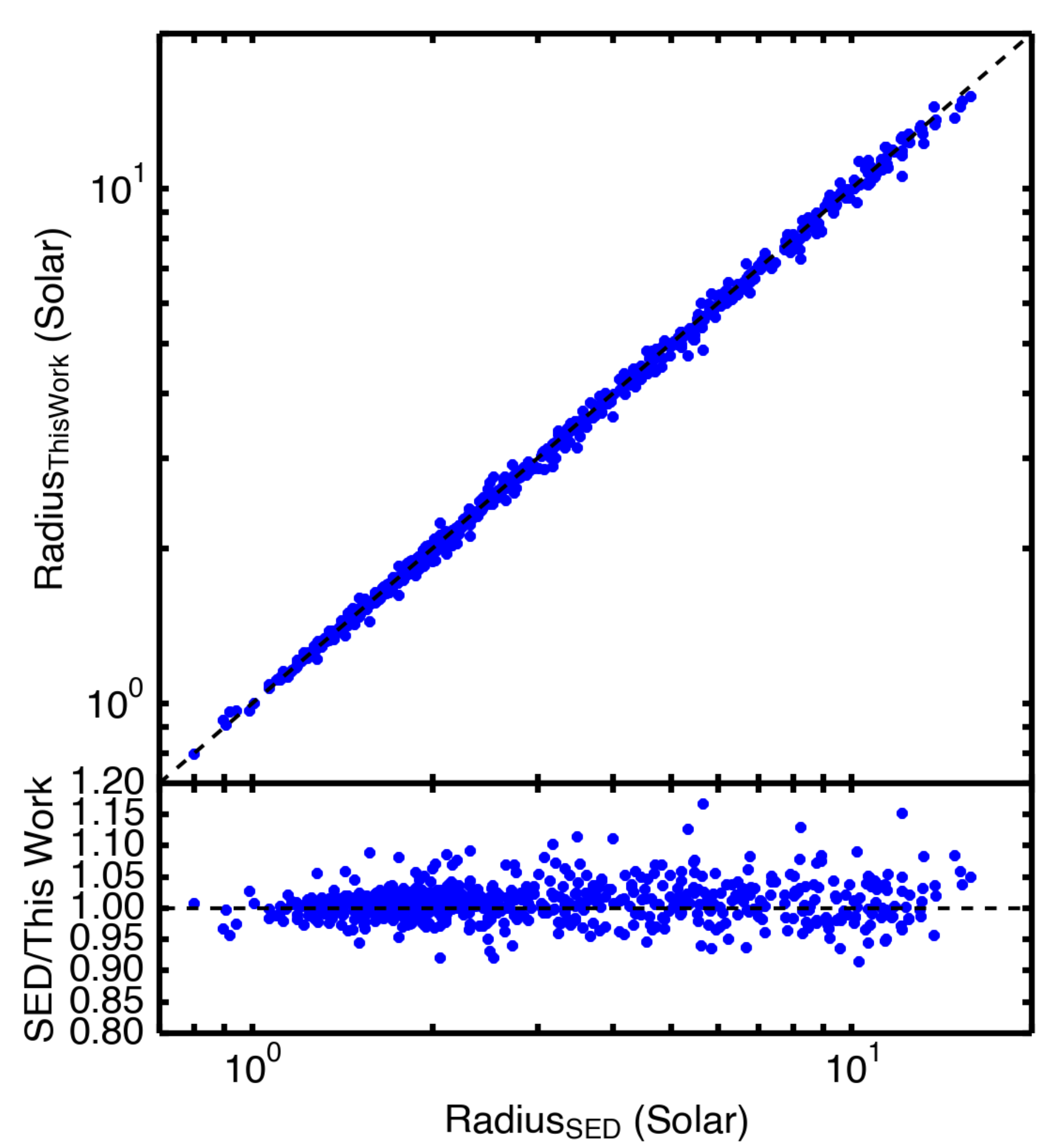}}
\caption{Comparison between radii calculated from TGAS parallaxes and bolometric corrections adopted from the MIST/C3K grid versus bolometric fluxes measured using SED fitting as described in \citet{stassun16a}. The black dashed line shows the 1:1 relation.}
\label{fig:sed}
\end{center}
\end{figure}

To test the effect of systematic errors in bolometric corrections, we used the method by \citet{stassun16a} to calculate bolometric fluxes by fitting spectral energy distributions to broadband photometry supplemented with a grid of ATLAS model atmospheres \citep{kurucz93}. The SED fits used the same \teff, \logg\ and \feh\ values as input constraints, but reddening was left as free parameter. We then used these bolometric fluxes with \teff\ to calculate angular diameters which, combined with TGAS parallaxes, resulted in a set of stellar radii that could be directly compared with the radii calculated from TGAS parallaxes and bolometric corrections (see Section 3.1). 
Figure \ref{fig:sed} shows a comparison between the two estimates. We observe good agreement, with a median difference of 0.7\% and a scatter of $\approx$\,3\%, and a small systematic trend with SED radii being larger by $\approx$\,1\% for red giants ($\approx$\,3--10\,$\rsun$). 

Since the MIST grid also uses ATLAS models, the above exercise is mostly sensitive to differences in deriving bolometric fluxes rather than systematic differences in model atmospheres. We therefore performed a second test by comparing distances calculated using the same seismic luminosity and reddening but bolometric corrections calculated from MARCS model atmospheres \citep{gustafsson08} provided by \citet{casagrande14b}, as implemented in BASTA (see also left panel of Figure 2). We observed an offset of $\approx$\,1\% (with distances calculated using MARCS bolometric corrections being larger), which was approximately constant in distance. Based on these two tests, we conclude that systematic errors due to bolometric corrections are at the $\approx$\,1\% level in radius and distance, which is small compared to the random uncertainties of TGAS parallaxes (see Figure \ref{fig:sample}).

\subsection{Code Availability}

The stellar classification software tools described above as well as all data to reproduce the results of this paper (Tables 1 \& 2) are publicly available at \url{https://github.com/danxhuber/isoclassify} \citep{isoclassify}. The tools can be used to derive posterior distributions for stellar parameters and distances given any input combination of asteroseismic, astrometric, photometric and spectroscopic observables.

\section{Results}

\subsection{Parallax Comparison}

\begin{figure}
\begin{center}
\resizebox{\hsize}{!}{\includegraphics{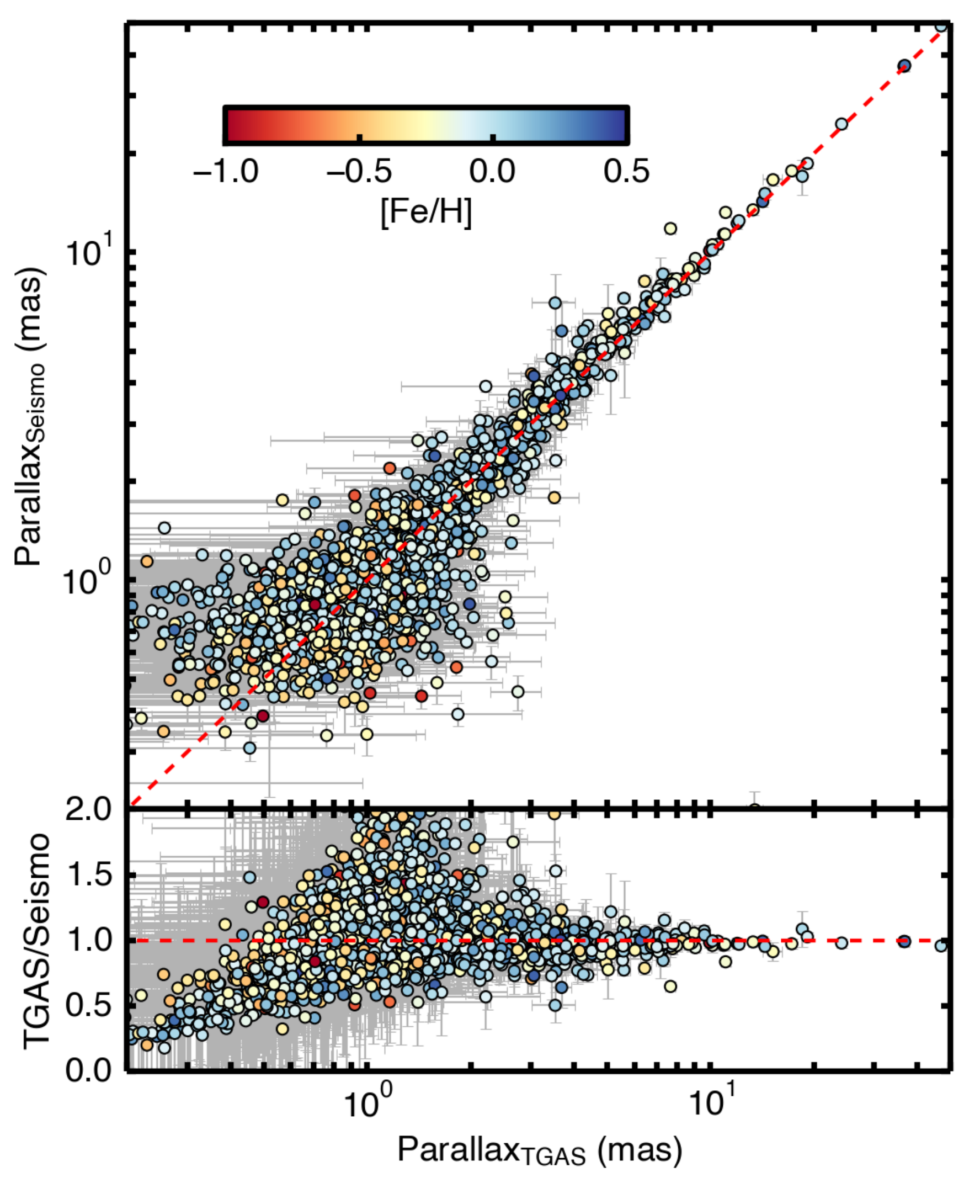}}
\caption{Asteroseismic parallaxes (calculated using the direct method without \Dnu\ correction) versus TGAS parallaxes for all 2200 stars in our sample. Metallicities are color-coded, and the dashed red line shows the 1:1 relation.}
\label{fig:parallax}
\end{center}
\end{figure}

Figure \ref{fig:parallax} compares parallaxes from asteroseismology with those from TGAS for all \nstars\ stars in our sample. We show results without \Dnu\ correction applied, but note that the effects of this correction are small compared to the scatter (see Section 4.2). Qualitatively, the comparison shows good agreement over three orders of magnitude. The scatter is dominated by large TGAS uncertainties for distant, evolved stars, which cause a diagonal ``edge'' in the ratios (bottom panel) toward low parallax values due to TGAS data systematically scattering to lower values than asteroseismology. This is mainly caused by asteroseismic distances being an order of magnitude more precise: because the giant sample is magnitude limited, we observe a lack of small parallax values from asteroseismology.

\begin{figure}
\begin{center}
\resizebox{\hsize}{!}{\includegraphics{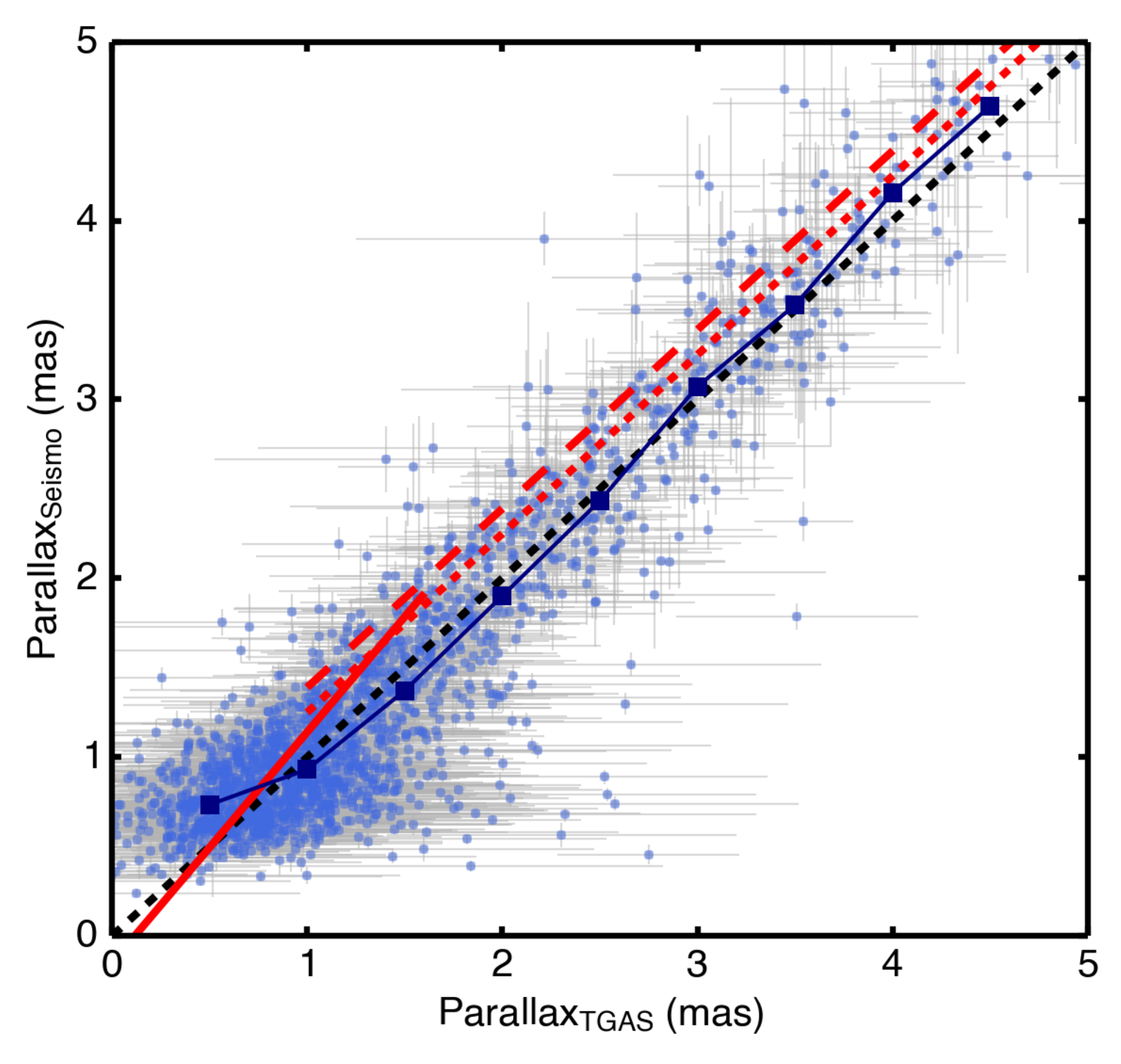}}
\caption{Asteroseismic versus TGAS parallaxes for stars with $\pi <$\,5\,mas. The dashed black line shows the 1:1 relation. Light blue symbols are individual stars, while thick dark blue squares show median bins spaced by 0.5\,mas. The red dashed and dotted lines show the predicted offsets from the TGAS parallax corrections by \citet{stassun16b} with and without ecliptic latitude dependence, respectively. The solid red line shows the predicted offset from the TGAS parallax correction by \citet{davies17}.}
\label{fig:zoom}
\end{center}
\end{figure}

The qualitative agreement in Figure \ref{fig:parallax} appears to contradict \citet{deridder16}, who reported that asteroseismic and TGAS parallaxes are incompatible with a 1:1 relation for $\approx$\,900 giants from \citet{rodrigues14}. To investigate this, we compare stars with parallaxes $<5$\,mas \citep[corresponding roughly to the largest parallax in the sample by][]{rodrigues14} on a linear scale in Figure \ref{fig:zoom}. We indeed observe a deviation from the 1:1 relation, with seismic parallaxes being systematically larger. However, the larger sample used here, which covers the transition from red giants to main sequence stars (Figure 1), demonstrates that this deviation appears to be significantly smaller than previously thought. Specifically, the TGAS parallax corrections derived from eclipsing binaries by \citet{stassun16b}, which indicated that TGAS parallaxes are too small ($\pi_{\rm{TGAS-EB}}=-0.25$\,mas using the mean offset or $\pi_{\rm{TGAS-EB}}=-0.39$\,mas using an ecliptic latitude $\beta=55$\,degrees) are significantly too large. There is also tension with the upper end of the \citet{davies17} correction (which predicts a similar offset to \citet{stassun16b} at $\approx$\,1.6\,mas). We note these results are not significantly affected by the small offset between our distances and \citet{rodrigues14} discussed in Section 3.3.1.

In agreement with the combined results by \citet{sesar16}, \citet{jao16} and \citet{davies17} we find that the absolute offset increases for larger parallaxes, which on average correspond to less evolved stars. This implies a stronger absolute systematic offset for main-sequence stars and subgiants, which is surprising given that scaling relations are generally thought to be more reliable for stars similar to the Sun. However, asteroseismic distances scale as $\teff^{2.5}$, which varies significantly for main-sequence and subgiant stars. Indeed, \teff\ scales are often plagued by systematic offsets \citep[e.g.][]{pinsonneault11}. In general, photometric $\teff$ scales from the infrared flux method \citep{casagrande11} or open clusters \citep{an13} are systematically hotter than spectroscopic temperatures, although recent color-$\teff$ calibrations are consistent with or cooler than spectroscopy \citep{huang15b}. All $\teff$ scales rely on the accuracy of interferometric angular diameters \citep[e.g.][]{boyajian12,boyajian12b,white13}, some of which have been suspected to be affected by systematic errors \citep{casagrande14c}. While efforts to systematically cross-calibrate angular diameters between different instruments are currently underway \citep[e.g.][]{huber16b}, it is still unclear which $\teff$ scale is indeed most accurate.

\begin{figure*}
\begin{center}
\resizebox{\hsize}{!}{\includegraphics{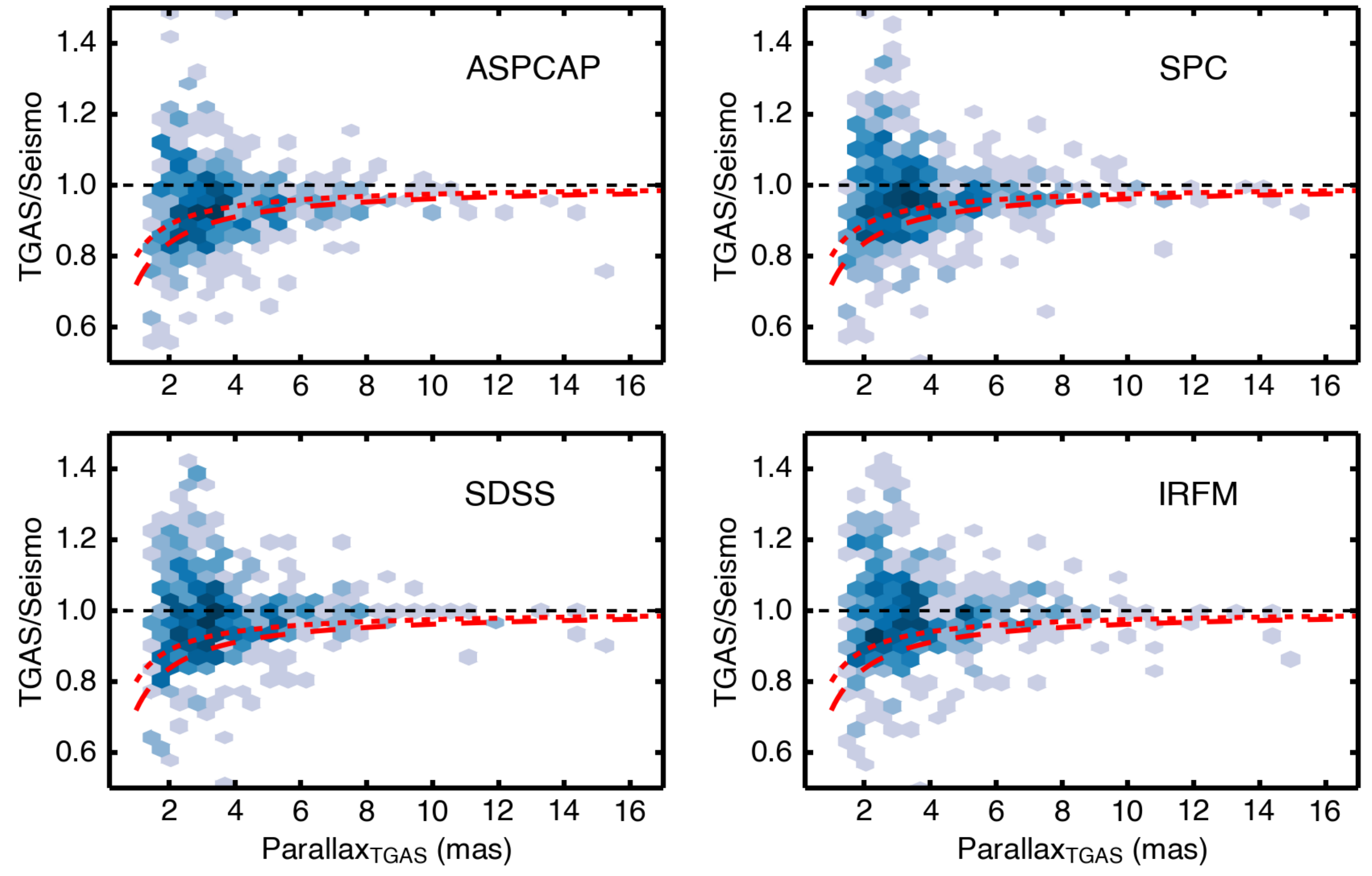}}
\caption{Ratio of asteroseismic and TGAS parallaxes as a function of TGAS parallax for the dwarf and subgiant sample with $\pi > 1.5$\,mas. Colors show the logarithmic number density, with darker colors corresponding to a higher number of stars. Each panel shows a different adopted \teff\ scale to calculate asteroseismic parallaxes. The average temperature offsets are $\Delta(\teff)_{\rm SDSS-ASPCAP}\approx 220$\,K, $\Delta(\teff)_{\rm IRFM-ASPCAP}\approx 270$\,K, $\Delta(\teff)_{\rm SDSS-SPC}\approx 110$\,K and $\Delta(\teff)_{\rm IRFM-SPC}\approx 140$\,K. We note that ASPCAP temperatures are not calibrated for dwarfs \citep{holtzman15}, and hence are likely underestimated. The red dashed and dotted lines show the predicted offsets from the TGAS parallax corrections by \citet{stassun16b} with and without ecliptic latitude dependence, respectively.}
\label{fig:teffcalib}
\end{center}
\end{figure*}

To test the effect of changing the \teff\ scale, we recalculated asteroseismic distances for dwarfs and subgiants using temperatures from the APOGEE pipeline (ASPCAP), and also using photometric \teff\ values from the infrared flux method \citep[IRFM, ][]{casagrande11} and Sloan photometry \citep[SDSS, ][]{pinsonneault11} as listed in \citet{pinsonneault11}. We note that that \citet{pinsonneault11} used $\feh=-0.2$\,dex and extinction values from the KIC, which were shown to be overestimated compared to values derived from asteroseismology and spectroscopy \citep{rodrigues14}. Accounting for these differences would result in shifts of $\approx -20$\,K for the SDSS and $\approx -65$\,K for the IRFM scales, depending on the adopted initial \teff\ and extinctions. Furthermore, the SDSS and IRFM scales are not entirely independent, since SDSS was calibrated to match IRFM for $\teff > 6000$\,K. Re-deriving the SDSS and IRFM \teff\ scales for the sample is beyond the scope of this paper, but we note that neither of these effects significantly change the conclusions below.

For the comparison, we discarded stars with $\pi < 1.5$\,mas to avoid the ``edge'' bias that arises from large uncertainty differences discussed above. The average difference between the coolest (ASPCAP) and hottest (IRFM) \teff\ scale is $\approx$\,270\,K. The results in Figure \ref{fig:teffcalib} demonstrate that the hotter \teff\ scales bring better agreement between asteroseismic and TGAS parallaxes, particularly for $\pi \lesssim$\,10\,mas. Specifically, the median offset over the whole sample reduces by more than a factor of 2 from $5.8\pm0.6$\% for the coolest \teff\ scale (ASPCAP) to $2.0\pm0.7$\% for the IRFM. Figure \ref{fig:teffcalib} also shows the proposed corrections by \citet{stassun16b} derived from eclipsing binaries. The $-0.25$\,mas correction, which was the main result of the study, provides a good match to the data for $\pi \gtrsim $\,5\,mas and spectroscopic \teff\ scales, but is overestimated for $\pi \lesssim 5$\,mas for all \teff\ scales. The correction including an ecliptic latitude dependence is overestimated for $\pi \lesssim 10$\,mas for all \teff\ scales.

In summary, our analysis demonstrates that offsets between TGAS parallaxes, asteroseismology and eclipsing binaries are likely smaller than previously reported for $\pi \lesssim 5-10$\,mas ($\gtrsim$\,100--200\,pc), and can be at least partially compensated by systematic errors in \teff\ scales for dwarfs and subgiants. Residual differences are small fractions rather than absolute offsets, and are $\approx$\,2\,\% for the hottest \teff\ scales. This conclusion is consistent with \citet{silva17} and \citet{jao16}, who found agreement with the offset by \citet{stassun16b} for nearby dwarfs for which $\approx$\,2\,\% produces a $-0.25$\,mas offset. These results imply that previously proposed TGAS parallax corrections may be overestimated for $\pi \lesssim 5-10$\,mas ($\approx$\,90--98\% of the TGAS sample). We note that this difference is most likely due to the larger sample size used in this study, rather than systematic differences in the adopted methods or distance scales. The above results also provide empirical evidence that hotter \teff\ scales (such as the infrared flux method) are more accurate than cooler, spectroscopic estimates. Importantly, this conclusion assumes that there are no strong systematic errors in TGAS and asteroseismic distances.

\subsection{Radius Comparison}

\begin{figure}
\begin{center}
\resizebox{\hsize}{!}{\includegraphics{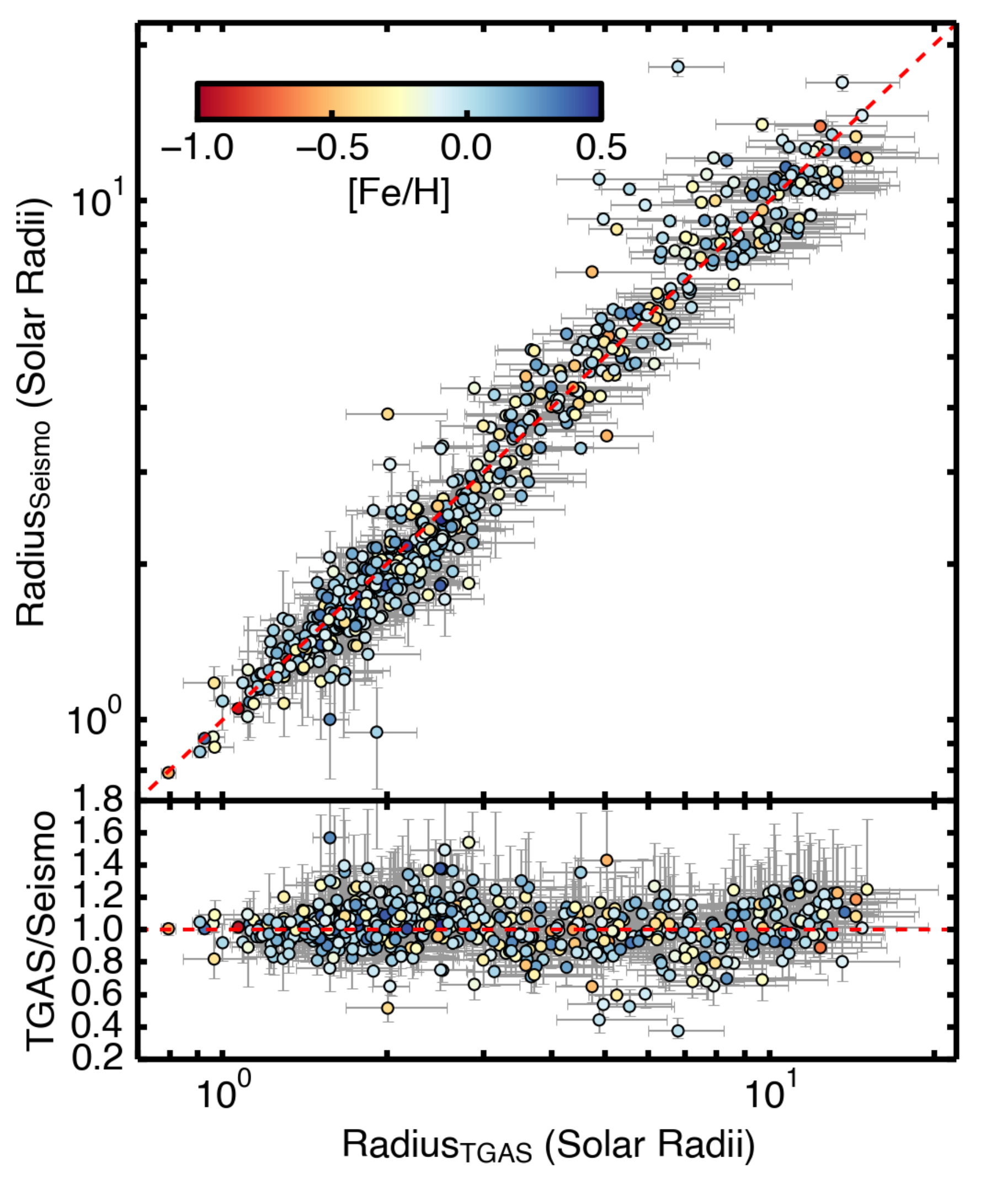}}
\caption{Comparison of radii derived using TGAS parallaxes with radii calculated from asteroseismic scaling relations for stars with $\sigma_{\pi}/\pi < 0.2$. No \Dnu\ correction was applied. Color-coding denotes the metallicity for each star. The average residual median and scatter is $\sim$\,3\% and $\sim$\,10\%, respectively.}
\label{fig:rad}
\end{center}
\end{figure}

\begin{figure*}
\begin{center}
\resizebox{\hsize}{!}{\includegraphics{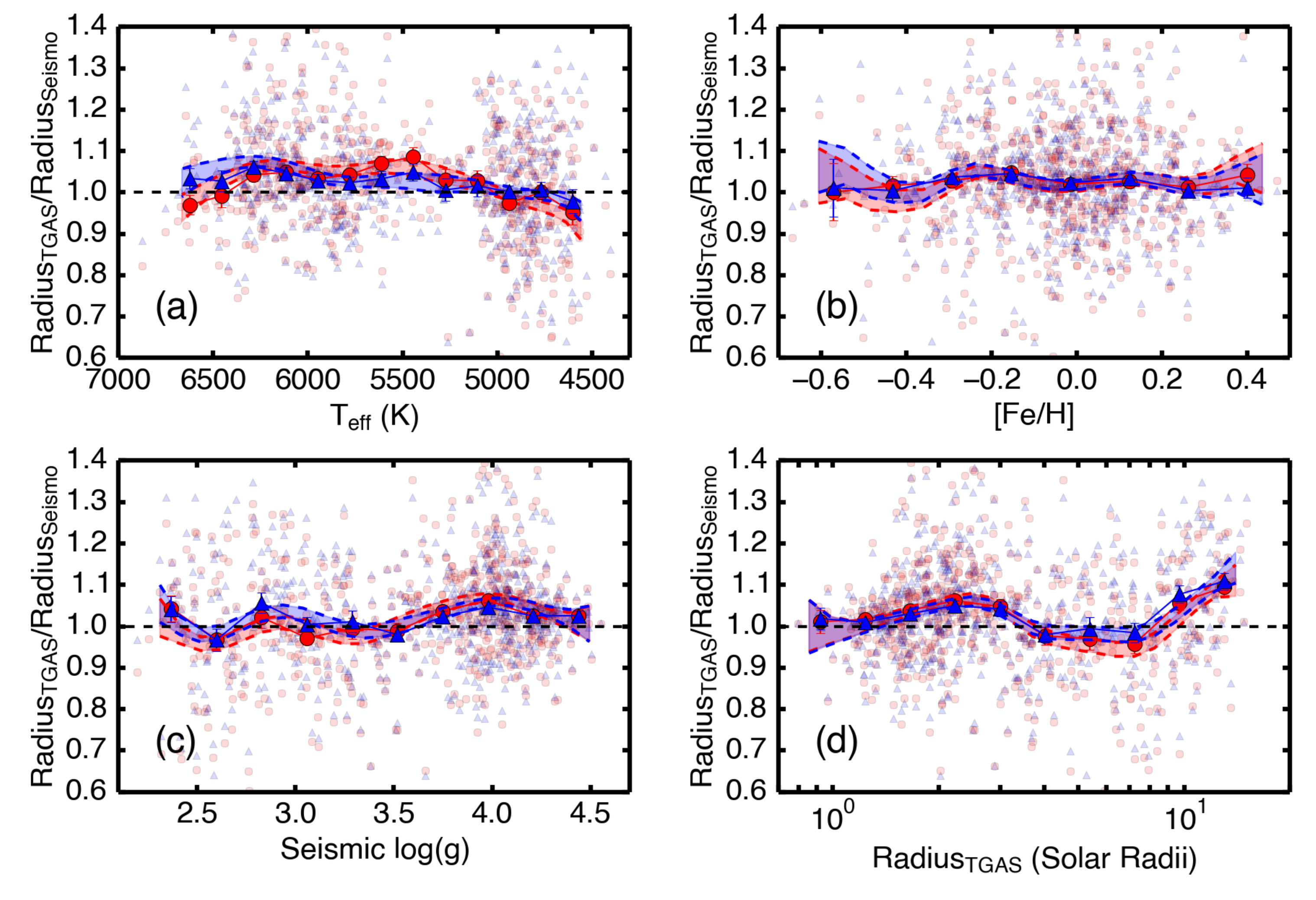}}
\caption{Ratio of TGAS radii over asteroseismic radii as a function of \teff, \logg\, \feh\ and TGAS radius. Small red circles and blue triangles show unbinned data with and without applying the \citet{sharma16} $\Dnu$ scaling relation correction, respectively. Thick symbols show median binned data. 
Shaded areas and dashed lines show 68\% confidence intervals calculated by bootstrapping a local-quadratic nonparametric regression using \texttt{pyqt-fit}.
Note that the upturn for large radii is an artifact due to the large uncertainty differences between both samples (see text and Figure \ref{fig:radcomp_all}).}
\label{fig:rad_residuals}
\end{center}
\end{figure*}

Comparing radii instead of parallaxes reduces the \teff\ dependence (from $\teff^{2.5}$ to $\teff^{1.5}$), and allows a more direct test of a fundamental parameter predicted by scaling relations. Figure \ref{fig:rad} compares asteroseismic and TGAS radii for all stars with a TGAS parallax measured to better than 20\%, which approximately corresponds to the limit where the distance ratios are not heavily influenced by the exponentially decreasing volume density prior (see Section 3.1) or artefacts introduced by large differences in random errors (see Section 4.1). The overall agreement is excellent, empirically demonstrating that asteroseismic radii from scaling relations without any corrections are accurate to at least $\approx$\,10\,\% for stars ranging from $\approx 0.8$ to $10\,\rsun$. The color-coding in Figure \ref{fig:rad} furthermore demonstrates that there are no strong biases in asteroseismic radii as a function of metallicity.

To illustrate this further, Figure \ref{fig:rad_residuals} shows the ratios as a function of $\teff$, $\logg$, $\feh$ and TGAS radius, both with and without applying the \Dnu\ correction by \citet{sharma16}. In addition to the raw data (small symbols) we also show median bins (large symbols). We have tested that spatial correlations between asteroseismic and TGAS parallaxes (Zinn et al., in prep) do not significantly affect these median values or their uncertainties for the typical spatial separations of stars contributing to a given bin ($\approx$\,1.5 degrees). We also show 68\% confidence intervals calculated by bootstrapping a local-quadratic nonparametric regression using \texttt{pyqt-fit}\footnote{\url{http://pyqt-fit.readthedocs.io/en/latest/modules.html}}.

We observe no significant trends with metallicity for $\feh = -0.8$ to $+0.4$ dex (Figure \ref{fig:rad_residuals}b). Intriguingly, however, the ratios show a trend as a function of TGAS radius (Figure \ref{fig:rad_residuals}d): stars near the main sequence ($\sim$1--1.5$\rsun$) show no offset, while the seismic radii of subgiants ($\sim$1.5--3$\rsun$) are too small by $\approx$\,5--7\%. The offset reduces for low-luminosity red giants, before increasing for high-luminosity red giants ($\gtrsim 10\rsun$). The \Dnu\ scaling relation correction slightly reduces these deviations (blue triangles). The upturn for high-luminosity red giants ($\gtrsim 10\rsun$) in Figure \ref{fig:rad_residuals}d is artificially introduced by large uncertainties of TGAS radii in a magnitude-limited sample, similar to the ``edge'' bias for parallaxes in the bottom panel of Figure \ref{fig:parallax}. The underestimated seismic radii for subgiants, however, cannot be explained by such an effect. 

\begin{figure}
\begin{center}
\resizebox{\hsize}{!}{\includegraphics{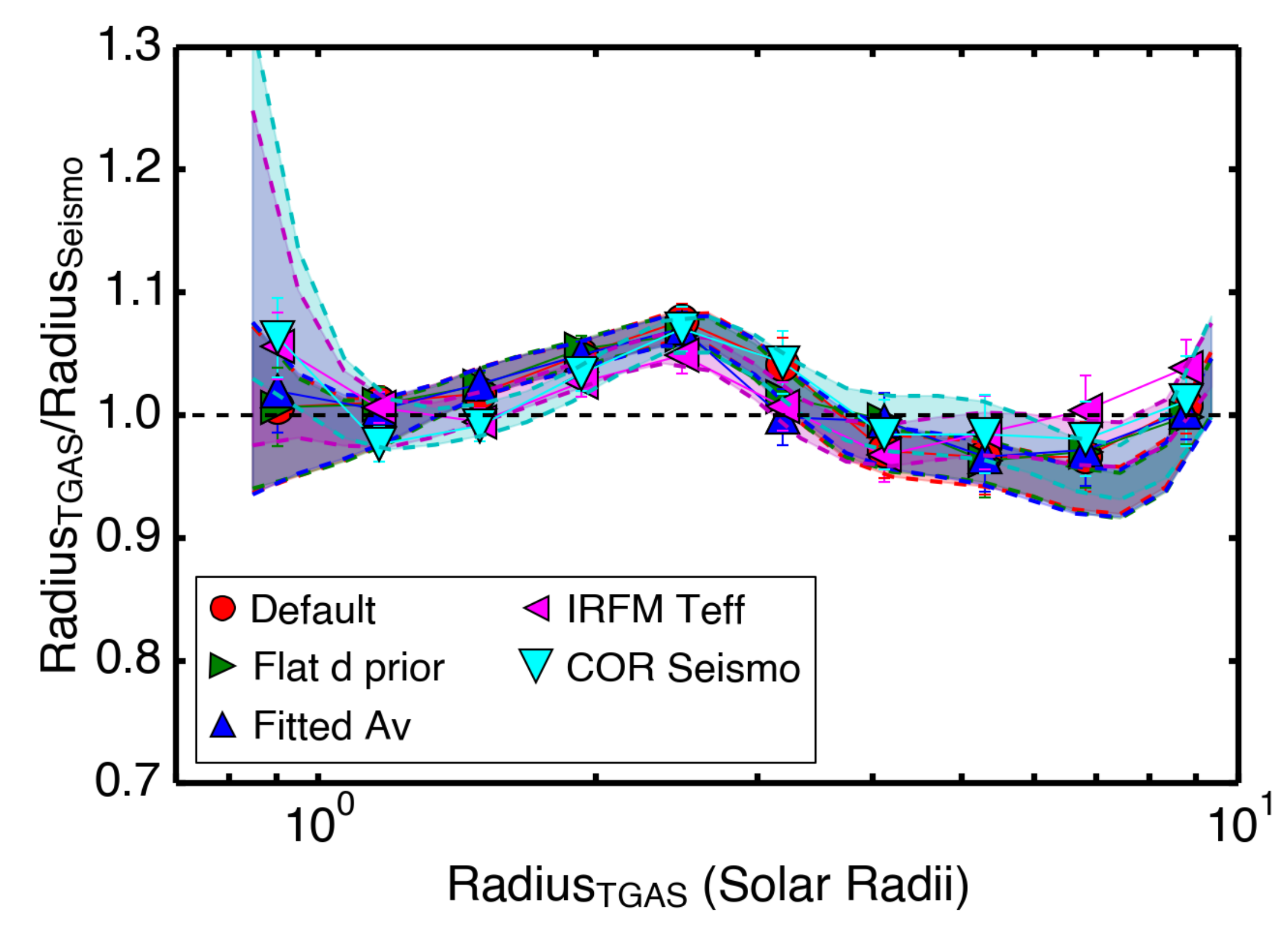}}
\caption{Same as Figure \ref{fig:rad_residuals}d but restricting the sample to stars with $R<10\rsun$ and using no \Dnu\ correction (red circles). Different symbols and colors show the same analysis repeated assuming a flat distance prior (green right-facing triangles), using reddening values measured using grid-modeling (blue upwards triangles), using IRFM temperatures (magenta left-facing triangles), and using seismic parameters from the COR pipeline (cyan downwards triangles).Large uncertainties at the lowest radii are caused by the sparseness of cool dwarfs in some of the test samples.}
\label{fig:radcomp_allb}
\end{center}
\end{figure}

We confirmed that the radius trend in Figure \ref{fig:rad_residuals}d is independent of the distance prior, reddening, method for calculating asteroseismic observables, or adopted \teff\ scales (see Figure \ref{fig:radcomp_allb}). Note that we have excluded giants with $R>10\rsun$ from this comparison to remove the bias discussed above. Specifically, we used a flat distance prior, reddening measured from the grid-modeling method described in Section 3.2, as well as $\numax$ and $\Dnu$ values from the COR pipeline \citep{mosser09}. Adopting IRFM \teff\ values for dwarfs and subgiants instead of the default SPC scale reduced the offset for subgiants by $\approx$\,2\,\% (magenta symbols Figure \ref{fig:radcomp_allb}). We added this value in quadrature in the subsequent analysis to account for \teff-dependent systematics.

\begin{figure}
\begin{center}
\resizebox{\hsize}{!}{\includegraphics{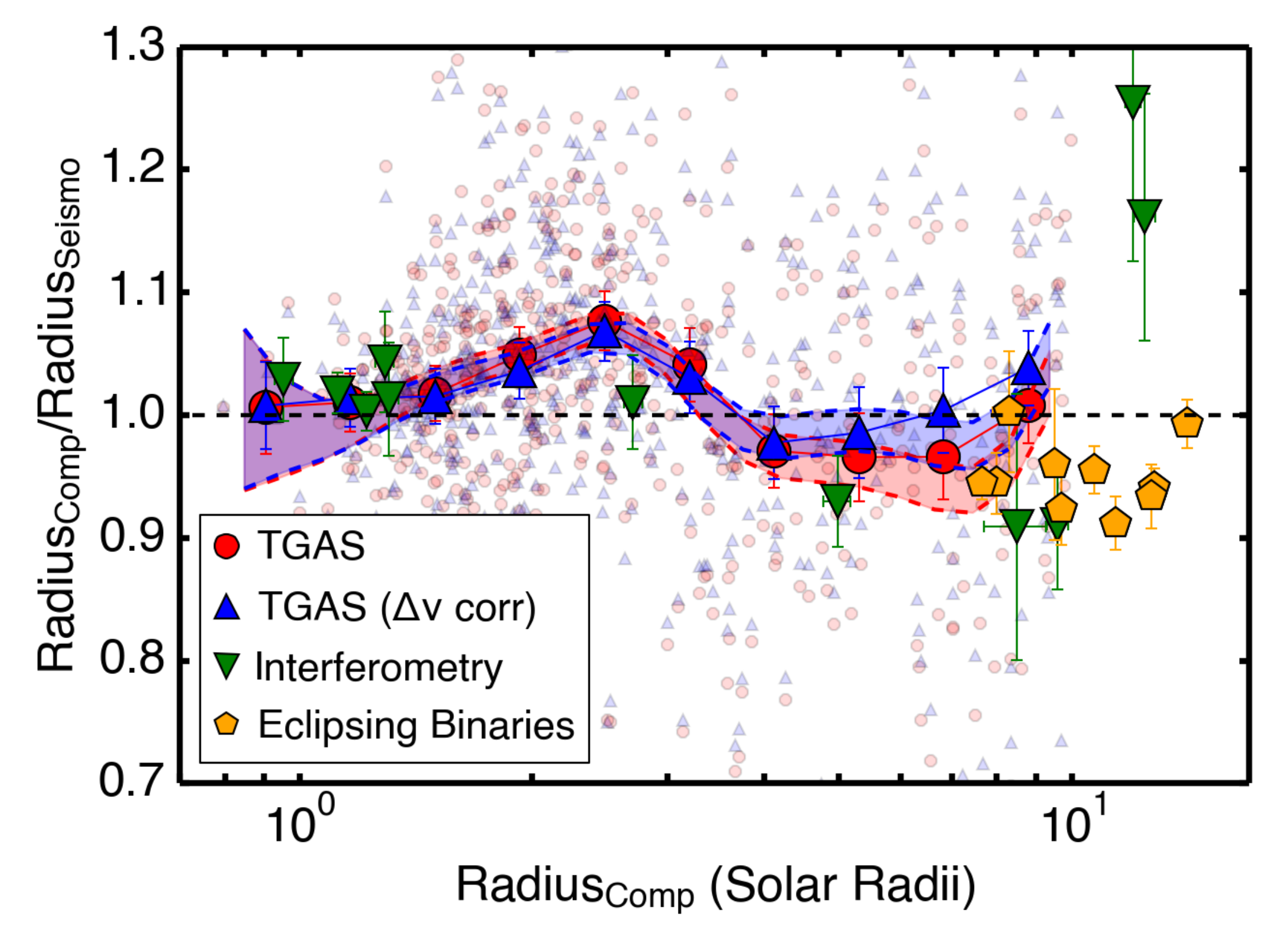}}
\caption{Comparison of asteroseismic radii derived from scaling relations with radii derived from four methods. Red circles and blue upward triangles show our TGAS sample with and without the \citet{sharma16} \Dnu\ scaling relation correction, and shaded areas show 68\% confidence intervals as in Figure \ref{fig:rad_residuals}. We also show stars with interferometrically measured radii  \citep[green triangles,][]{huber12,white13,johnson14} and red giants in double-lined eclipsing binary systems \citep[orange pentagons,][]{gaulme16}.}
\label{fig:radcomp_all}
\end{center}
\end{figure}

To put the TGAS radius comparison into context, Figure \ref{fig:radcomp_all} also shows results from eclipsing binaries \citep{gaulme16} and interferometry \citep{huber12,white13,johnson14}. The interferometry sample is sparse for subgiants, but does not strongly contradict the $\approx$\,5\,\% bias for subgiants from TGAS. For giants our results are compatible with \citet{gaulme16}, although the $\Dnu$-corrected results are in slight tension with their predicted 5\,\% offset. Either way, the TGAS results imply that the $\approx$5\,\% radius bias reported by \citet{gaulme16} does not seem to extend the regime of low-luminosity red giants, which are prime targets for studies of exoplanets orbiting asteroseismic hosts \citep{grunblatt16}. A larger interferometric sample (White et al., in prep) as well as spectrophotometric angular diameters in combination with Gaia parallaxes (Grunblatt et al., in prep) will allow us to confirm and quantify the trends in Figure \ref{fig:radcomp_all}. Table \ref{tab:ratios} lists the median binned ratios shown in Figure \ref{fig:radcomp_all}, which may be used to estimate systematic errors in seismic radii from scaling relations.

\setcounter{table}{2}
\begin{table}
\begin{small}
\begin{center}
\caption{Median Binned Ratios between Gaia and Seismic Radii}
%\vspace{0.1cm}
\begin{tabular}{c c c}        
\hline         
$R_{\rm Gaia} (\rsun)$ & $R_{\rm Gaia}/R_{\rm seismo}$ &  $R_{\rm Gaia}/R_{\rm seismo, \Dnu corr}$ \\
\hline  
  0.90 &  1.007 $\pm$  0.038 &  1.008 $\pm$  0.035   \\ 
  1.16 &  1.010 $\pm$  0.024 &  1.014 $\pm$  0.024   \\ 
  1.50 &  1.018 $\pm$  0.023 &  1.015 $\pm$  0.022   \\ 
  1.93 &  1.049 $\pm$  0.023 &  1.036 $\pm$  0.023   \\ 
  2.48 &  1.077 $\pm$  0.024 &  1.069 $\pm$  0.024   \\ 
  3.20 &  1.041 $\pm$  0.030 &  1.031 $\pm$  0.029   \\ 
  4.12 &  0.971 $\pm$  0.030 &  0.977 $\pm$  0.029   \\ 
  5.30 &  0.966 $\pm$  0.036 &  0.986 $\pm$  0.037   \\ 
  6.83 &  0.966 $\pm$  0.035 &  1.004 $\pm$  0.035   \\ 
  8.79 &  1.008 $\pm$  0.030 &  1.039 $\pm$  0.030   \\ 
\hline
\end{tabular} 
\label{tab:ratios} 
\end{center}
\flushleft Note: $R_{\rm seismo, \Dnu corr}$ corresponds to seismic radii derived using the $\Dnu$ scaling relation correction by \citet{sharma16} (i.e. blue symbols in Figure \ref{fig:radcomp_all}). Uncertainties include a 2\% systematic error due to different \teff\ scales.
\end{small}
\end{table}

\subsection{Red-Giant Branch versus Red Clump}

Models of red giants lead us to expect a systematic difference in the $\Dnu$ scaling relation as a function of the evolutionary state due to the changes in their interior sound-speed profile after the onset of He-core burning \citep{miglio12c}. However, the degree and even the sign of this difference is not yet fully settled. For example, \citet{miglio12c} showed that applying the \Dnu\ correction to red clump stars improves the agreement with independent radii measured in clusters, while the results by \citet{sharma16} implied that the largest effect of the \Dnu\ correction applies for ascending RGB stars. Previous samples to empirically test scaling relations have been too small to decide this question.

\begin{figure}
\begin{center}
\resizebox{\hsize}{!}{\includegraphics{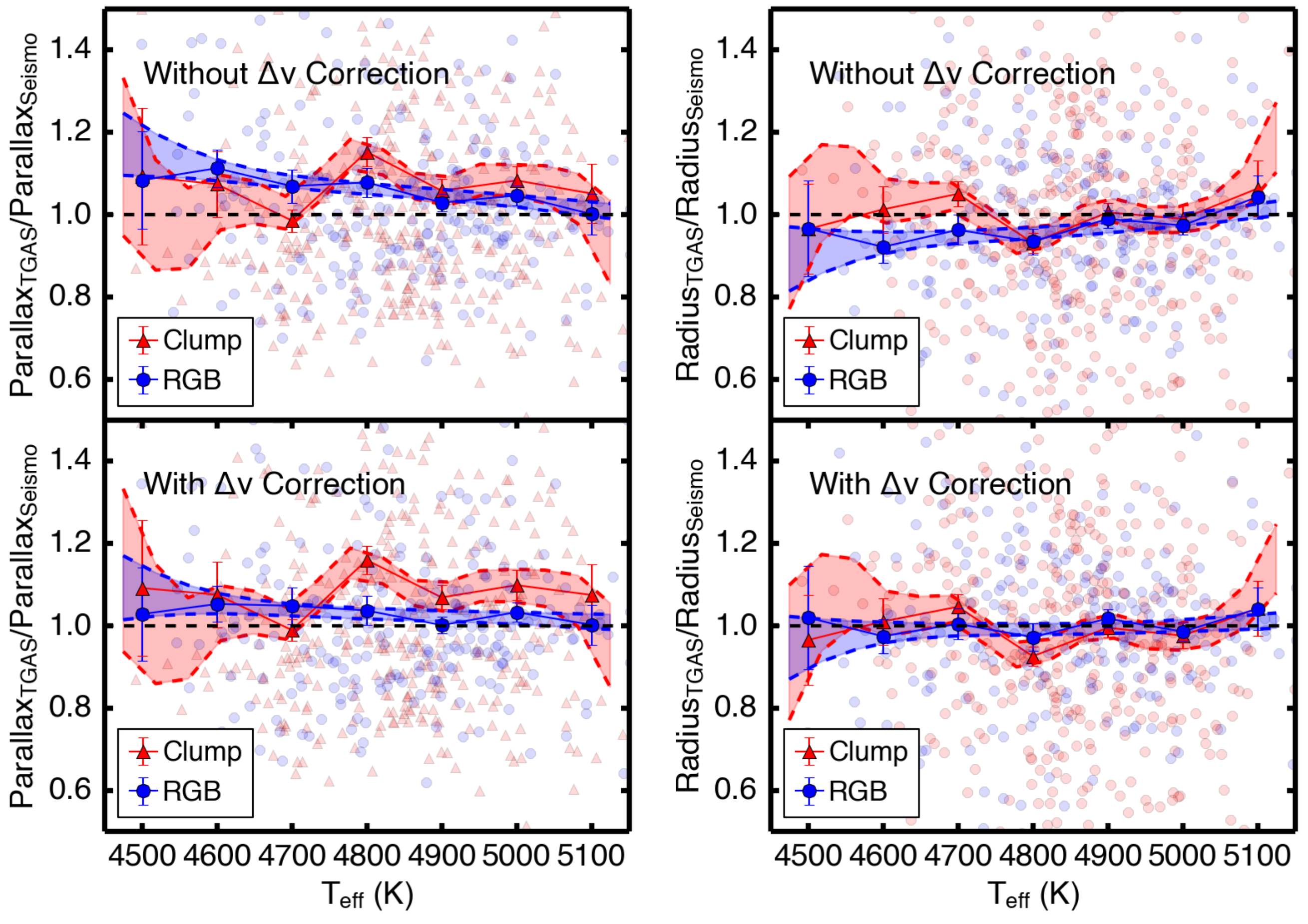}}
%\resizebox{\hsize}{!}{\includegraphics{rgbclump_1.pdf}}
%\resizebox{\hsize}{!}{\includegraphics{rgbclump_2.pdf}}
\caption{Comparison of parallaxes (left panels) and radii (right panels) for ascending RGB (blue circles) and red clump (red triangles) stars, respecively. Top panels show the comparison without applying a correction to the \Dnu\ scaling relation, while the bottom panels show the comparison with the \citet{sharma16} correction applied. Small symbols show the original sample, large symbols with error bars are mean bins, and shaded areas show 68\% confidence intervals as in Figure \ref{fig:rad_residuals}.}
\label{fig:rgbclump}
\end{center}
\end{figure}

TGAS parallaxes allow us to test the dependency of the scaling relation correction on evolutionary state. To separate RGB and red clump stars, we used classifications based on mixed mode period spacings by \citet{stello13} and \citet{vrard16}. Figure \ref{fig:rgbclump} shows parallaxes (left panels) and radii (right panels) both with (bottom) and without (top) applying the \Dnu\ scaling relation correction by \citet{sharma16}. The samples in each panel are separated into RGB (blue circles) and red clump stars (red triangles). Note that we relaxed the fractional parallax uncertainty cut to $<$\,40\% to include more red clump stars in the sample. Due to this relaxed cut the median bins were offset from the local-quadratic fit, and we thus adopted mean bins for consistency. However, the conclusions below are not unaffected by whether mean or median bins are used.

While the scatter is too large to determine whether the RGB or red clump stars agree better with TGAS, there is tentative evidence that the \Dnu\ correction provides a improvement for RGB stars. Specifically, the weighted mean offset reduces from $5.4\pm1.3$\% to $2.7\pm0.7$\% in parallax and from $-3.1\pm1.4$\% to $-1.0\pm 1.5$\% for radius. The corrections for red clump stars are negligible, as expected. We conclude that TGAS parallaxes are not precise enough to decide how the \Dnu\ correction depends on evolutionary state, but provide tentative evidence (at the $\approx$\,2-$\sigma$ level)  that the \citet{sharma16} corrections improve the accuracy of seismic distances and radii.

%\subsection{Which Scaling Relation is more reliable?}

%\begin{figure}
%\begin{center}
%\resizebox{\hsize}{!}{\includegraphics{dnunumaxcomp.pdf}}
%\caption{Comparison of parallaxes (left panels) and radii (right panels) for ascending RGB (blue circles) and red clump (red triangles) stars, respecively. Top panels show the comparison without applying a correction to the \Dnu\ scaling relation, while the bottom panels show the comparison with the \citet{sharma16} correction applied. Small symbols show the original sample, large symbols with error bars show binned distributions.}
%\label{fig:rgbclump}
%\end{center}
%\end{figure}

\subsection{Synergies of Gaia and Asteroseismic Distances}

TGAS provides a first glimpse of the potential of Gaia to measure distances, and vast precision improvements are expected for upcoming data releases. Since asteroseismic and TGAS distances agree to within a few percent over several orders of magnitude, it is interesting to explore the complementary nature of Gaia and asteroseismology to measure distances to galactic stellar populations. To investigate this, we calculated the expected end-of-mission Gaia parallax precision for seismic \kep\ targets using the Gaia performance model\footnote{\url{http://www.cosmos.esa.int/web/gaia/science-performance}}:

\begin{equation}
\begin{split}
\sigma_{\pi}/\rm{mas} = \sqrt{(-1.631+680.766 z + 32.732 z^{2})} \\
 (0.986 + (1 - 0.986) V-I_{C}) \: ,
\end{split}
\label{equ:gaia}
\end{equation}

\noindent
with

\begin{equation}
z =
\begin{cases}
    0.0685                          & \text{for } G < 12.1\\
    10^{0.4 (G-15)}              & \text{otherwise } 
\end{cases}
\label{equ:class1}
\end{equation}

Here, $\sigma_{\pi}$ is the predicted end-of-mission parallax uncertainty averaged over the sky. The Gaia $G$-band magnitude and Johnson-Cousins $V-I$ color were calculated from KIC $gri$ photometry (Table 1) using the following relations \citep{jordi06,jordi10}:

\begin{equation}
\begin{split}
G = (-0.0662 - 0.7854 (g-r) - 0.2859 (g-r)^{2} + \\ 
0.0145 (g-r)^{3}) + g \: ,
\end{split}
\end{equation}

\noindent
and

\begin{equation}
V-I =
\begin{cases}
0.675 (g-r) + 0.364\, & \text{for } g-r < 2.1   \\
1.11 (g-r) - 0.52\, &    \text{otherwise }  \: .
\end{cases}
\end{equation}

To account for the sky-position dependency of parallax uncertainties due to the Gaia scanning law, we interpolated the recommended scaling factors\footnote{\url{http://www.cosmos.esa.int/web/gaia/table-6}} for the ecliptic coordinates of each \kep\ target. This yielded on average $\sim$28\% smaller uncertainty than the uncertainties calculated from Equation \ref{equ:gaia}. 

\begin{figure}
\begin{center}
\resizebox{\hsize}{!}{\includegraphics{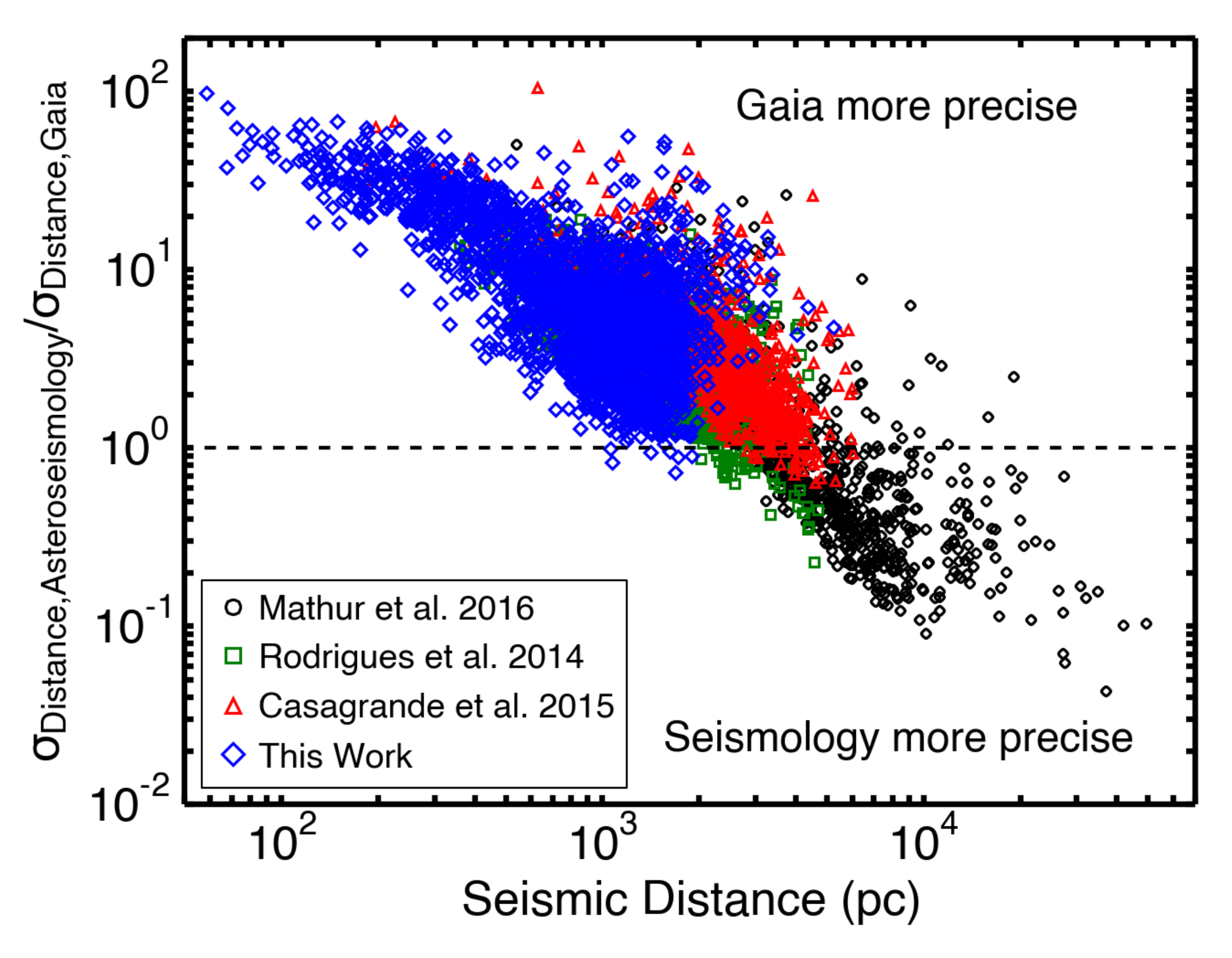}}
\caption{Asteroseismic distance precision divided by the expected end-of-mission Gaia uncertainty as a function of seismic distance for different samples of \kep\ targets. Note that the sample in this work and by \citet{casagrande14} contains dwarfs and giants, while \citet{rodrigues14} and \citet{mathur16} analyzed giants only.}
\label{fig:distcomp}
\end{center}
\end{figure}

Figure \ref{fig:distcomp} compares the distance uncertainty from asteroseismology to the expected end-of-mission Gaia precision for stars with asteroseismic distances from this work, \citet{rodrigues14}, \citet{casagrande14} and \citet{mathur16}\footnote{Note that \citet{mathur16} did not include uncertainties due to extinction, which however are not expected to dominate the error budget: e.g. $\sigma_{A_{J}}=0.03$\,mag corresponds to a $\approx$\,1\,\% error in distance, which is much smaller than the typical $\approx$\,5\,\% distance uncertainty in \citet{mathur16}.}.
Remarkably, asteroseismology will provide more precise distances than the best Gaia performance for stars beyond 3\,kpc. This is because the asteroseismic sensitivity does not  depend strongly on apparent magnitude and hence distant, high luminosity red giants still yield precisions of a few percent out to tens of kpc \citep{mathur16}. Asteroseismology will therefore be critical to extend the reach of Gaia to distant stellar populations, particularly if combined with spectroscopy, which simultaneously allows to constrain interstellar extinction (Figure \ref{fig:discomp}). Current and future opportunities to detect oscillations in distant red giants outside the \kep\ field include the K2 Mission \citep{howell14}, targets with 1-year coverage near the ecliptic poles observed by TESS \citep{ricker14}, red giants in the bulge observed with WFIRST \citep{gould15}, and red giants observed with PLATO \citep{rauer14}.

\section{Conclusions}

We presented a detailed comparison of asteroseismic scaling relations with Gaia DR1 (TGAS) parallaxes for \nstars\ \kep\ stars spanning from the main sequence to the red-giant branch. Our main findings can be summarized as follows:

\begin{itemize}
\item Previously identified offsets between TGAS parallaxes and distances derived from asteroseismology and eclipsing binaries have likely been overestimated for stars beyond 100--200\,pc in the \kep\ field. This implies that previously proposed TGAS parallax corrections are likely overestimated for $\pi \lesssim 5-10$\,mas ($\approx$\,90--98\% of the TGAS sample). We emphasize that this is most likely due to the larger sample size used here, rather than systematic differences in the methods or distance scales in previous studies. We demonstrate that for subgiants and dwarfs the offsets can be in part compensated by adopting a hotter $\teff$ scale (such as the infrared flux method) as opposed to spectroscopic temperatures. If systematics from scaling relations and TGAS parallaxes are negligible, these results would validate the IRFM as a fundamental $\teff$ scale for dwarfs and subgiants. Residual systematic differences between asteroseismology and TGAS parallaxes are a constant fraction (at the $\approx$\,2\,\% level) across three orders of magnitude, in line with the previously noted dependence of absolute TGAS parallax offsets with distance. 

\item Asteroseismic and Gaia radii agree with a residual scatter of $\approx$\,10\% but reveal a systematic offset for subgiants ($\approx$1.5--3\rsun), with seismic radii being underestimated by $\approx$\,5--7\%, with a $\approx$\,2\,\% systematic error depending on the \teff\ scale. Our results show no significant offsets for main-sequence stars ($\lesssim 1.5\rsun$) and low-luminosity giants with ($R \approx$\,3--8\rsun), indicating that the offsets derived from eclipsing binaries by \citet{gaulme16} do not appear to extend to less evolved stars. Overall, our results demonstrate empirically that systematic errors in radii derived from scaling relations are at or below the $\approx$\,5\% level from $\approx 0.8-10\,\rsun$.

\item A comparison of parallaxes and radii for RGB and red clump stars shows tentative evidence (at the $\approx$\,2\,$\sigma$ level) that the $\Dnu$ scaling relation correction by \citet{sharma16} improves the comparison to Gaia. However, the precision of TGAS parallaxes is insufficient to conclusively show whether the \Dnu\ correction is more important for RGB or red clump stars.

\item Our results provide no evidence for systematic errors in asteroseismic scaling relations as a function of metallicity from $\feh \approx -0.8$ to $+0.4$\,dex. This provides empirical support for the use of asteroseismology to calibrate spectroscopic pipelines for characterizing exoplanet host stars \citep[e.g.][]{brewer15} and galactic archeology \citep[e.g.][]{valentini16}.

\item We used the Gaia performance model to predict that asteroseismic distances will remain more precise than Gaia end-of-mission data for stars beyond $\approx$\,3\,kpc. This highlights the complementary nature of Gaia and asteroseismology for measuring distances to galactic stellar populations.
\end{itemize}

The study presented here only gives a first glimpse of the powerful synergy between Gaia and asteroseismology. In-depth studies using individual frequency modeling using TGAS parallaxes will provide further insights into differences in distance scales and seismic fundamental parameters \citep[e.g.][]{metcalfe17}, and new interferometry as well as spectrophotometry for dozens of seismic red giants will provide a more fundamental calibration of the scaling relation for stellar radii. Furthermore, Gaia DR2 is expected to provide parallaxes for nearly all $\approx$\,20,000 oscillating \kep\ stars \citep[e.g.][]{mathur17}, allowing unprecedented scaling relation tests and studies which can combine frequency modeling and Gaia data to test and improve interior models from the main sequence to the red-giant branch.

\section*{Acknowledgments}
We thank Willie Torres, Yvonne Elsworth and our anonymous referee for helpful comments and discussions, as well as the entire \kep\ and \textit{Gaia} teams for making this paper possible. DH acknowledges support by the Australian Research Council's Discovery Projects funding scheme (project number DE140101364) and support by the National Aeronautics and Space Administration under Grant NNX14AB92G issued through the Kepler Participating Scientist Program. AS is partially supported by grant ESP2015-66134-R (MINECO).
VSA acknowledges support from VILLUM FONDEN (research grant 10118).
WJC and GRD acknowledge support from the UK Science and Technology Facilities Council. RAG acknowledges the support of CNES. Funding for the Stellar Astrophysics Centre is provided by The Danish National Research Foundation (Grant agreement no.: DNRF106).
SM acknowledges support from NASA grants NNX12AE17G, NNX15AF13G, and NNX14AB92G, as well as NSF grant AST-1411685.  

This work has made use of data from the European Space Agency (ESA) mission {\it Gaia} (\url{https://www.cosmos.esa.int/gaia}), processed by the {\it Gaia} Data Processing and Analysis Consortium (DPAC, \url{https://www.cosmos.esa.int/web/gaia/dpac/consortium}). Funding for the DPAC has been provided by national institutions, in particular the institutions participating in the {\it Gaia} Multilateral Agreement. Funding for the \kep\ Mission is provided by NASA's Science Mission Directorate. 
This publication makes use of data products from the Two Micron All Sky Survey, which is a joint project of 
the University of Massachusetts and the Infrared Processing and Analysis Center/California Institute of Technology, 
funded by the National Aeronautics and Space Administration and the National Science Foundation.
Funding for the Sloan Digital Sky Survey IV has been provided by the
Alfred P. Sloan Foundation, the U.S. Department of Energy Office of
Science, and the Participating Institutions. SDSS acknowledges
support and resources from the Center for High-Performance Computing at
the University of Utah. The SDSS web site is www.sdss.org. SDSS is managed by the Astrophysical Research Consortium for the Participating Institutions of the SDSS Collaboration including the Brazilian Participation Group, the Carnegie Institution for Science, Carnegie Mellon University, the Chilean Participation Group, the French Participation Group, Harvard-Smithsonian Center for Astrophysics, Instituto de Astrofísica de Canarias, The Johns Hopkins University, Kavli Institute for the Physics and Mathematics of the Universe (IPMU) / University of Tokyo, Lawrence Berkeley National Laboratory, Leibniz Institut für Astrophysik Potsdam (AIP), Max-Planck-Institut für Astronomie (MPIA Heidelberg), Max-Planck-Institut für Astrophysik (MPA Garching), Max-Planck-Institut für Extraterrestrische Physik (MPE), National Astronomical Observatories of China, New Mexico State University, New York University, University of Notre Dame, Observatório Nacional / MCTI, The Ohio State University, Pennsylvania State University, Shanghai Astronomical Observatory, United Kingdom Participation Group, Universidad Nacional Autónoma de México, University of Arizona, University of Colorado Boulder, University of Oxford, University of Portsmouth, University of Utah, University of Virginia, University of Washington, University of Wisconsin, Vanderbilt University, and Yale University.

\setcounter{table}{0}
%\begin{table*}
\begin{sidewaystable}
\begin{tiny}
\begin{center}
\caption{Observational Data}
\begin{tabular}{c c c c c c c c c c c c c}
KIC & $\numax (\muHz)$ & $\Dnu (\muHz)$ & $\pi$ (mas) & $B_{T}$ (mag) & $V_{T}$ (mag) & $g$ (mag) & $r$ (mag) & $i$ (mag) & $z$ (mag) & $J$ (mag) & $H$ (mag) & $K$ (mag) \\
\hline
   1160789 & 25.221 $\pm$ 0.760 &  3.529 $\pm$ 0.063 &  1.350 $\pm$ 0.322 & 11.159 $\pm$ 0.051 & 10.059 $\pm$ 0.031 & 10.418 $\pm$ 0.020 &  9.635 $\pm$ 0.020 &  9.342 $\pm$ 0.020 &  9.195 $\pm$ 0.020 &  8.133 $\pm$ 0.021 &  7.593 $\pm$ 0.021 &  7.497 $\pm$ 0.017  \\ 
   1161618 & 34.363 $\pm$ 0.599 &  4.100 $\pm$ 0.029 &  1.228 $\pm$ 0.358 & 11.905 $\pm$ 0.101 & 10.657 $\pm$ 0.054 & 11.052 $\pm$ 0.020 & 10.138 $\pm$ 0.020 &  9.818 $\pm$ 0.020 &  9.652 $\pm$ 0.020 &  8.542 $\pm$ 0.018 &  7.974 $\pm$ 0.026 &  7.887 $\pm$ 0.018  \\ 
   1162746 & 28.042 $\pm$ 1.268 &  3.710 $\pm$ 0.128 &  0.952 $\pm$ 0.981 & 13.455 $\pm$ 0.392 & 11.815 $\pm$ 0.165 & 12.203 $\pm$ 0.020 & 11.409 $\pm$ 0.020 & 11.075 $\pm$ 0.020 & 10.878 $\pm$ 0.020 &  9.834 $\pm$ 0.022 &  9.272 $\pm$ 0.020 &  9.183 $\pm$ 0.018  \\ 
   1163621 & 51.170 $\pm$ 0.863 &  5.005 $\pm$ 0.026 &  0.787 $\pm$ 0.386 & 13.144 $\pm$ 0.331 & 12.044 $\pm$ 0.190 & 12.597 $\pm$ 0.020 & 11.731 $\pm$ 0.020 & 11.401 $\pm$ 0.020 & 11.188 $\pm$ 0.020 & 10.107 $\pm$ 0.022 &  9.558 $\pm$ 0.018 &  9.473 $\pm$ 0.018  \\ 
   1294385 & 106.498 $\pm$ 1.084 &  9.113 $\pm$ 0.015 &  1.425 $\pm$ 0.515 & 12.177 $\pm$ 0.156 & 11.027 $\pm$ 0.076 & 11.363 $\pm$ 0.020 & 10.574 $\pm$ 0.020 & 10.296 $\pm$ 0.020 & 10.139 $\pm$ 0.020 &  9.085 $\pm$ 0.018 &  8.595 $\pm$ 0.021 &  8.465 $\pm$ 0.018  \\ 
   1430163 & 1775.247 $\pm$ 72.128 & 85.873 $\pm$ 1.879 &  5.486 $\pm$ 0.352 & 10.159 $\pm$ 0.027 &  9.627 $\pm$ 0.023 &  9.694 $\pm$ 0.020 &  9.480 $\pm$ 0.020 &  9.429 $\pm$ 0.020 &  9.459 $\pm$ 0.020 &  8.769 $\pm$ 0.026 &  8.560 $\pm$ 0.018 &  8.529 $\pm$ 0.018  \\ 
   1433803 & 150.146 $\pm$ 0.997 & 12.179 $\pm$ 0.017 &  2.683 $\pm$ 0.341 & 11.833 $\pm$ 0.089 & 10.503 $\pm$ 0.046 & 10.967 $\pm$ 0.020 & 10.090 $\pm$ 0.020 &  9.814 $\pm$ 0.020 &  9.630 $\pm$ 0.020 &  8.538 $\pm$ 0.020 &  8.046 $\pm$ 0.020 &  7.942 $\pm$ 0.018  \\ 
   1435467 & 1382.311 $\pm$ 9.148 & 70.558 $\pm$ 0.053 &  5.598 $\pm$ 0.249 &  9.484 $\pm$ 0.020 &  9.017 $\pm$ 0.016 &  9.021 $\pm$ 0.020 &  8.778 $\pm$ 0.020 &  8.685 $\pm$ 0.020 &  8.666 $\pm$ 0.020 &  7.983 $\pm$ 0.024 &  7.753 $\pm$ 0.023 &  7.718 $\pm$ 0.017  \\ 
   1435573 & 25.220 $\pm$ 0.773 &  3.728 $\pm$ 0.091 &  0.636 $\pm$ 0.323 & 13.664 $\pm$ 0.422 & 11.880 $\pm$ 0.175 & 12.277 $\pm$ 0.020 & 11.338 $\pm$ 0.020 & 11.004 $\pm$ 0.020 & 10.787 $\pm$ 0.020 &  9.690 $\pm$ 0.021 &  9.150 $\pm$ 0.022 &  8.991 $\pm$ 0.018  \\ 
   1569842 & 134.456 $\pm$ 0.659 & 11.765 $\pm$ 0.017 &  1.246 $\pm$ 0.562 & 11.933 $\pm$ 0.098 & 11.163 $\pm$ 0.079 & 11.785 $\pm$ 0.020 & 11.010 $\pm$ 0.020 & 10.736 $\pm$ 0.020 & 10.582 $\pm$ 0.020 &  9.589 $\pm$ 0.022 &  9.074 $\pm$ 0.016 &  8.989 $\pm$ 0.018  \\ 
 
\hline
\end{tabular}
\label{tab:observables}
\flushleft Notes: \numax\ and \Dnu\ were calculated using the SYD pipeline \citet{huber09} and taken from version 3.6.5 of the APOKASC catalog (Pinsonneault et al., in prep) for giants and from a reanalysis of the \citet{chaplin14} sample for dwarfs and subgiants (Serenelli et al., in prep). Note that for our analysis we added a 1\% and 0.5\% uncertainty in \numax and \Dnu\ to the formal uncertainties listed here to account for differences between asteroseismic analysis methods. $gri$ denotes KIC photometry converted into the SDSS scale using the transformations by \citet{pinsonneault11}. 
\end{center}
\end{tiny}
%\end{table*}
\end{sidewaystable}

\newcolumntype{H}{>{\setbox0=\hbox\bgroup}c<{\egroup}@{}}

\setcounter{table}{1}
%\begin{table*}
\begin{sidewaystable}
\begin{tiny}
\begin{center}
\caption{Derived Fundamental Properties, Distances and Extinctions}
\begin{tabular}{c | c c c | c c c | c c c | c c c | HHHHHHHHH}
KIC & \multicolumn{3}{|c|}{Spectroscopy + Asteroseismology} & \multicolumn{3}{c|}{Direct Method} & \multicolumn{3}{c|}{Direct Method ($\Dnu$ corr)} & \multicolumn{3}{|c|}{Grid Modeling} &  &  &  &  & \\
\hline
    & \teff (K)  & \logg & \feh & R(\rsun) & d (pc) & $A_{V}$ & R(\rsun) & d (pc) & $A_{V}$ & R(\rsun) & d (pc) & $A_{V}$ & R(\rsun) & d (pc) & $A_{V}$ & -- & --  \\ 
\hline
   1160789 &  4739 $\pm$  86 &  2.307 $\pm$ 0.014 & -0.340 $\pm$ 0.060 & 10.84$^{+0.54}_{-0.54}$ &   656$^{+42}_{-42}$ & 0.246 & 9.85$^{+0.49}_{-0.49}$ &   597$^{+38}_{-38}$ & 0.227 & 11.29$^{+0.38}_{-0.31}$ &   677$^{+24}_{-19}$ & 0.133$^{+0.070}_{-0.070}$ & 11.29$^{+0.34}_{-0.31}$ &   679$^{+25}_{-21}$ & 0.133$^{+0.070}_{-0.070}$ & 11.92$^{+7.10}_{-2.13}$ &   754$^{+405}_{-147}$ & 0.545 &  -1 &     apo      \\ 
   1161618 &  4763 $\pm$  86 &  2.442 $\pm$ 0.009 & -0.009 $\pm$ 0.060 & 10.96$^{+0.31}_{-0.31}$ &   806$^{+39}_{-39}$ & 0.085 & 11.01$^{+0.31}_{-0.31}$ &   810$^{+39}_{-39}$ & 0.086 & 11.21$^{+0.08}_{-0.07}$ &   800$^{+ 6}_{- 6}$ & 0.223$^{+0.080}_{-0.070}$ & 11.21$^{+0.09}_{-0.07}$ &   801$^{+ 7}_{- 6}$ & 0.223$^{+0.070}_{-0.070}$ & 11.58$^{+12.53}_{-2.68}$ &   808$^{+881}_{-176}$ & 0.108 &  1 &     apo      \\ 
   1162746 &  4798 $\pm$  86 &  2.356 $\pm$ 0.020 & -0.478 $\pm$ 0.060 & 10.97$^{+0.92}_{-0.92}$ &  1452$^{+135}_{-135}$ & 0.251 & 10.88$^{+0.92}_{-0.92}$ &  1441$^{+134}_{-134}$ & 0.250 & 11.02$^{+0.80}_{-0.58}$ &  1435$^{+114}_{-76}$ & 0.353$^{+0.080}_{-0.090}$ & 11.02$^{+0.80}_{-0.58}$ &  1435$^{+114}_{-76}$ & 0.353$^{+0.080}_{-0.080}$ & 7.52$^{+44.45}_{-2.62}$ &   919$^{+5788}_{-399}$ & 0.282 &  1 &     apo      \\ 
   1163621 &  4959 $\pm$  86 &  2.624 $\pm$ 0.008 & -0.041 $\pm$ 0.060 & 11.18$^{+0.29}_{-0.29}$ &  1722$^{+79}_{-79}$ & 0.372 & 11.34$^{+0.29}_{-0.29}$ &  1747$^{+80}_{-80}$ & 0.372 & 11.10$^{+0.18}_{-0.23}$ &  1676$^{+39}_{-34}$ & 0.383$^{+0.080}_{-0.070}$ & 11.14$^{+0.20}_{-0.15}$ &  1685$^{+34}_{-27}$ & 0.373$^{+0.080}_{-0.080}$ & 8.30$^{+28.31}_{-3.03}$ &  1216$^{+4349}_{-348}$ & 0.385 &  1 &     apo      \\ 
   1294385 &  4825 $\pm$  86 &  2.936 $\pm$ 0.006 &  0.030 $\pm$ 0.060 & 6.92$^{+0.14}_{-0.14}$ &   667$^{+29}_{-29}$ & 0.192 & 6.71$^{+0.13}_{-0.13}$ &   646$^{+28}_{-28}$ & 0.189 & 6.93$^{+0.16}_{-0.12}$ &   668$^{+17}_{-12}$ & 0.073$^{+0.070}_{-0.070}$ & 6.58$^{+0.32}_{-0.12}$ &   633$^{+31}_{-11}$ & 0.043$^{+0.100}_{-0.090}$ & 7.81$^{+27.34}_{-2.19}$ &   754$^{+2612}_{-209}$ & 0.287 &  -1 &     apo      \\ 
   1430163 &  6590 $\pm$  77 &  4.226 $\pm$ 0.018 & -0.050 $\pm$ 0.101 & 1.52$^{+0.09}_{-0.09}$ &   181$^{+12}_{-12}$ & 0.057 & 1.47$^{+0.09}_{-0.09}$ &   176$^{+12}_{-12}$ & 0.056 & 1.48$^{+0.03}_{-0.03}$ &   175$^{+ 4}_{- 4}$ & 0.103$^{+0.070}_{-0.070}$ & 1.46$^{+0.03}_{-0.02}$ &   174$^{+ 4}_{- 4}$ & 0.103$^{+0.070}_{-0.070}$ & 1.53$^{+0.11}_{-0.10}$ &   182$^{+13}_{-11}$ & 0.057 &  0 &     spc      \\ 
   1433803 &  4721 $\pm$  86 &  3.081 $\pm$ 0.005 &  0.198 $\pm$ 0.060 & 5.41$^{+0.10}_{-0.10}$ &   405$^{+18}_{-18}$ & 0.152 & 5.23$^{+0.10}_{-0.10}$ &   392$^{+17}_{-17}$ & 0.149 & 5.38$^{+0.14}_{-0.09}$ &   403$^{+10}_{- 6}$ & 0.083$^{+0.080}_{-0.080}$ & 5.14$^{+0.09}_{-0.07}$ &   384$^{+ 7}_{- 6}$ & 0.053$^{+0.080}_{-0.080}$ & 4.97$^{+0.84}_{-0.59}$ &   375$^{+60}_{-40}$ & 0.147 &  0 &     apo      \\ 
   1435467 &  6326 $\pm$  77 &  4.108 $\pm$ 0.004 &  0.010 $\pm$ 0.101 & 1.72$^{+0.03}_{-0.03}$ &   138$^{+ 5}_{- 5}$ & 0.068 & 1.69$^{+0.03}_{-0.03}$ &   136$^{+ 5}_{- 5}$ & 0.067 & 1.70$^{+0.02}_{-0.02}$ &   137$^{+ 2}_{- 2}$ & 0.003$^{+0.070}_{-0.070}$ & 1.68$^{+0.02}_{-0.01}$ &   135$^{+ 2}_{- 2}$ & 0.033$^{+0.070}_{-0.080}$ & 2.23$^{+0.12}_{-0.12}$ &   177$^{+ 9}_{- 7}$ & 0.092 &  0 &     spc      \\ 
   1435573 &  4678 $\pm$  86 &  2.304 $\pm$ 0.014 &  0.020 $\pm$ 0.060 & 9.64$^{+0.58}_{-0.58}$ &  1156$^{+83}_{-83}$ & 0.258 & 9.36$^{+0.56}_{-0.56}$ &  1122$^{+81}_{-81}$ & 0.255 & 11.09$^{+0.24}_{-0.20}$ &  1286$^{+28}_{-23}$ & 0.283$^{+0.080}_{-0.080}$ & 11.07$^{+0.22}_{-0.18}$ &  1284$^{+28}_{-20}$ & 0.283$^{+0.080}_{-0.080}$ & 12.98$^{+36.18}_{-4.02}$ &  1470$^{+4308}_{-345}$ & 0.316 &  -1 &     apo      \\ 
   1569842 &  4847 $\pm$  86 &  3.039 $\pm$ 0.004 & -0.317 $\pm$ 0.060 & 5.26$^{+0.09}_{-0.09}$ &   640$^{+27}_{-27}$ & 0.291 & 5.03$^{+0.09}_{-0.09}$ &   613$^{+26}_{-26}$ & 0.283 & 5.26$^{+0.04}_{-0.02}$ &   648$^{+ 6}_{- 4}$ & 0.153$^{+0.070}_{-0.070}$ & 5.02$^{+0.04}_{-0.06}$ &   618$^{+ 6}_{- 9}$ & 0.183$^{+0.070}_{-0.070}$ & 6.36$^{+34.96}_{-1.79}$ &   774$^{+4208}_{-216}$ & 0.416 &  0 &     apo      \\ 
 
\hline
\end{tabular}

\bigskip

\begin{tabular}{HHHHHHHHHHHHH | c c c | c c c | c c}
 &  &  & & & & & & & & & & & \multicolumn{3}{|c|}{Grid Modeling ($\Dnu$ corr)} & \multicolumn{3}{|c|}{Gaia Parallax} & Ev & Src \\
\hline
    & \teff (K)  & \logg & \feh & R(\rsun) & d (pc) & $A_{V}$ & R(\rsun) & d (pc) & $A_{V}$ & R(\rsun) & d (pc) & $A_{V}$ & R(\rsun) & d (pc) & $A_{V}$ & R(\rsun) & d (pc) & $A_{V}$ & -- & -- \\ 
\hline
 
\hline
\end{tabular}

\label{tab:results}
\flushleft Notes: \teff\ and \feh\ were taken from APOGEE DR13 \citep{sdssdr13} for giants (src flag = 'apo') and from \citet{buchhave14} for dwarfs and subgiants (src flag = 'spc'). \logg\ was calculated from \teff\ and \numax\ given in Table 1. 'Ev' denotes the evolutionary state for non He-core burning stars (0), He-core burning stars (1) and stars with unknown evolutionary state (-1). Evolutionary state classifications were taken from \citet{stello13} and \citet{vrard16}.
\end{center}
\end{tiny}
%\end{table*}
\end{sidewaystable}

\bibliographystyle{aasjournal}
\bibliography{../../tex/references}

\end{document}